\documentclass[12pt,letterpaper]{article}
 
\usepackage{amsmath}
\allowdisplaybreaks[1]

\usepackage{graphicx}
\usepackage{subfig}
\graphicspath{{fig/}}

\usepackage{amssymb}

\usepackage{rotating}
\usepackage{array}

\usepackage{listings}
\usepackage{color}
\definecolor{dkgreen}{rgb}{0,0.4,0}
\definecolor{gray}{rgb}{0.5,0.5,0.5}

\usepackage{longtable}

\def\be{\begin{equation}}
\def\ee{\end{equation}}
\def\ba{\begin{align}}
\def\ea{\end{align}}
\def\nn{\nonumber}

\def\half{{{1\over 2}}}

\newcommand{\eq}[1]{(\ref{#1})}

\newcommand{\comment}[1]{}
\def\eqdef{\overset{\text{def}}{=}}

\newcommand{\norder}[2][]{~\mathop{\vphantom{
\lvert}}\nolimits^{\circ}_{\circ}#2\mathop{\vphantom{
\rvert}}\nolimits^{\circ}_{\circ #1}~}

\newcommand{\bignorder}[2][]{~\mathop{\vphantom{
\lvert}}\nolimits^{\circ}_{\circ}#2\mathop{\vphantom{
\rvert}}\nolimits^{\circ}_{\circ #1}~}

\newcommand{\fln}{\left\lfloor n/2\right\rfloor}

\begin{document}


\begin{titlepage}
\vskip 1in
\begin{center}
{\Large
{Renormalization schemes for SFT solutions}}
\vskip 0.5in
{Joanna L. Karczmarek and Matheson Longton}
\vskip 0.3in
{\it 
Department of Physics and Astronomy\\
University of British Columbia
Vancouver, Canada}
\end{center}

\vskip 0.5in
\begin{abstract}
In this paper, we examine the space of renormalization schemes
compatible with the Kiermaier and Okawa \cite{Kiermaier:2007vu} framework for
constructing Open String Field Theory solutions based
on marginal operators with singular self-OPEs.  We show
that, due to freedom in defining the renormalization scheme
which tames these singular OPEs, the solutions obtained from the KO 
framework are not necessarily unique. We identify a multidimensional
space of SFT solutions corresponding to a single given marginal operator.
\end{abstract}
\end{titlepage}

\tableofcontents


\section{Introduction}
\label{sec:intro}

The problem of finding analytic String Field Theory (SFT) solutions 
corresponding to different boundary CFTs has
received plenty of attention in the last few years.
There are several different approaches to this problem,
including a perturbative approach based on marginal deformations of 
the boundary CFT, as well as a more general non-perturbative approach
based on boundary condition changing (bcc) operators.
In much of this work, analytic solutions to SFT are
constructed as wedge states with insertions.  
Operators inserted on the boundary of the wedge state
are built using either a marginal operator or a bcc operator, together
with ghosts and other universal elements such as the stress energy tensor.
The largest technical challenge to be overcome in constructing
SFT solutions is due to the generically singular nature of 
marginal deformation operators and bcc operators:
when these operators have singular OPEs, regularization
schemes must be introduced.
Earliest attempts to construct SFT solutions
dealt mainly with finite boundary operators (for example, 
\cite{Kiermaier:2007ba,Schnabl:2007az}),
but the challenge due to singular operators has been mostly overcome:
for marginal operators, in \cite{Kiermaier:2007vu,Kiermaier:2007ki},
and, for bcc operators, in several recent works, including
\cite{Kiermaier:2010cf,Maccaferri:2014cpa,Erler:2014eqa}.
Other approaches have also been considered in 
\cite{Fuchs:2007yy} and \cite{Inatomi:2012nv,Kishimoto:2013sra,Takahashi:2002ez}.

One issue that has not been addressed very much is that of
uniqueness of the SFT solutions being constructed.
In particular, there is the question of whether different
renormalization schemes could result in different SFT
solutions.
Here, we attempt to study the issue of uniqueness, focusing on
the formalism developed in \cite{Kiermaier:2007vu}.
This is one of the formalisms able to handle a time-dependent singular 
marginal deformation, such as  the time-symmetric
rolling tachyon solution generated by an exactly
marginal operator $\sqrt{2} \cosh (X^0/\sqrt{\alpha'})$.
Since the SFT solution is built out of renormalized operators,
we can ask whether different choices of renormalization schemes
will lead to different SFT solutions.

We do uncover a two-parameter family of SFT solutions all corresponding
to the same marginal operator and discuss the possibility
that more might exist.  We do not, at this point,
have a clear interpretation of these solutions.
Notice that previous numerical
studies of the rolling tachyon solutions with regular OPE have not 
always agreed on coefficients 
\cite{Kiermaier:2007ba,Kiermaier:2010cf,Coletti:2005zj}, 
and a few possible explanations for the meaning of those coefficients 
have been suggested in \cite{Ellwood:2007xr}. 
If these different solutions are gauge equivalent, our analytic approach might 
make it easier to demonstrate that fact.    Should the solutions, however,
prove not to be gauge equivalent, under the equivalence of
boundary CFTs and open SFT solutions each of them would correspond
to a new boundary CFT.

The structure of this paper is as follows: In section \ref{sec.rtc.ko} we review the conditions that
renormalized operators must meet for the construction of  \cite{Kiermaier:2007vu}
to be valid and provide a description of our initial approach.
In section \ref{sec.rtc.quadratic} we study the space of all possible renormalization schemes at
second order in the deformation parameter.  In section \ref{sec.rtc.cubic}
we consider the cubic order.  In section \ref{sec.rtc.all-orders} we consider all orders
for a particular two-parameter family of renormalization schemes and 
prove that all the conditions set out by \cite{Kiermaier:2007vu} 
are satisfied for this family.  In section \ref{sec.discussion}
we discuss which of the free parameters present in the
renormalization scheme actually affect the SFT solution and how.
Finally, we propose the existence of an even larger
family of solutions.  Appendix \ref{lemma} contains the proof of an important
technical result necessary to prove the first BRST condition
of \cite{Kiermaier:2007vu} which was not included in that paper.

\section{Setup} \label{sec.rtc.ko}

The approach taken in \cite{Kiermaier:2007vu} starts with
a marginal operator $V(t)$ with a self-OPE given by
\be
V(t)V(0)\sim\frac{1}{t^{2}}+O(1)
\label{eqn:selfOPE}
\ee
with no $\frac{1}{t}$ term.
A deformed boundary condition on the interval $(a,b)$ is
achieved by inserting an exponential of the marginal
operator integrated between $a$ and $b$,
defined in terms of a Taylor series in the deformation
parameter $\lambda$:
\be
e^{\lambda V(a,b)} = \sum_{n=0}^\infty \frac{\lambda^n}{n!} V(a,b)^{n}~,
\ee
where
\be
V(a,b)^{n} =\left(\int_{a}^{b}dt V(t)\right)^{n}~=
\int_{(a,b)^n} d^nt ~ V(t_1) \ldots V(t_n)~.
\ee
Since $V(t)$ has
a singular self-OPE, the the above expressions need to be regulated.
We will denote the regulated (or renormalized) operators by enclosing them 
with $[~~~]_r$.

In \cite{Kiermaier:2007vu}, a list of conditions which must be
satisfied by the renormalization procedure is given.
If these conditions are satisfied, the formal solution
constructed in \cite{Kiermaier:2007vu}
will satisfy the SFT equations of motion and be real;
however, different renormalization schemes can possibly lead
to different SFT solutions.  The main goal of this paper
is to examine the space of possible renormalization schemes 
compatible with the condition required for a real SFT solution.
We begin by reviewing the conditions that any
renormalization scheme must satisfy to construct a SFT
solution using the approach of \cite{Kiermaier:2007vu}.
These conditions are basically physical conditions which
ensure that when $[e^{\lambda V(a,b)}]_r$ is inserted
on the boundary, the effect is a conformal change of boundary
conditions on the interval $(a,b)$, and nothing else.

The first condition ensures that the insertion  $[e^{\lambda V(a,b)}]_r$ 
does not modify the boundary conditions away from the interval
$(a,b)$.  In particular, it requires that when products
of operators that are inserted away from each other are renormalized,
it is sufficient to renormalize each term separately.
In other words, the renormalized operator factorizes
for operators with disjoint support.  For example:
\begin{subequations}\label{eq.rtc.assume-all}
\be
\left[\ldots e^{\lambda_{1}V(a,b)}e^{\lambda_{2}V(c,d)}\ldots\right]_{r}=\left[\ldots e^{\lambda_{1}V(a,b)}\right]_{r}\left[e^{\lambda_{2}V(c,d)}\ldots\right]_{r},~\text{for}~ b<c~. \label{eq.rtc.assume-factor}
\ee
Further, changing the boundary condition on the 
interval $(a,b)$ and $(b,c)$ using the same deformation
parameter should be the same as changing the boundary
condition on the interval $(a,c)$.  In other words,
renormalization should not spoil factorization of 
exponentials.
\be
\left[\ldots e^{\lambda V(a,c)}\ldots\right]_{r}=\left[\ldots e^{\lambda V(a,b)}e^{\lambda V(b,c)}\ldots\right]_{r}~.\label{eq.rtc.assume-replace}
\ee
This condition was called the `replacement condition' in \cite{Kiermaier:2007vu} 
to differentiate it from the factorization condition \eq{eq.rtc.assume-factor}. We
we will continue to use this term. \\ The next two conditions ensure that the resulting boundary condition
is conformal.  The first condition defines two local (unintegrated) operators $O_L$ and $O_R$,
which  play an important role in the solution.  We have
\be
Q_{B}\left[e^{\lambda V(a,b)}\right]_{r}=\left[e^{\lambda V(a,b)}O_{R}(b)\right]_{r}-\left[O_{L}(a)e^{\lambda V(a,b)}\right]_{r}~,\label{eq.rtc.assume-Q1} 
\ee
which requires 
the existence and finiteness of the renormalized operators 
$\left[O_{L}(a)e^{\lambda V(a,b)}\right]_{r}$ and $\left[e^{\lambda V(a,b)}O_{R}(b)\right]_{r}$, implying that the
OPE of the marginal operator $V$ with $O_{L,R}$ is not so singular that  it cannot be renormalized within the scheme we choose. 
The second  of these two assumptions expresses the fact that $Q_B$ is anti-commuting:
\be
Q_{B}\left[O_{L}(a)e^{\lambda V(a,b)}\right]_{r}
=-\left[O_{L}(a)e^{\lambda V(a,b)}O_{R}(b)\right]_{r}~.\label{eq.rtc.assume-Q2}\\
\ee
To obtain a real solution, it is important not to violate the reflection 
symmetry:
\be
\left[e^{\lambda\int_{a}^{b}dt V(t)}\right]_{r}=\left[e^{\lambda\int_{a}^{b}dt V(a+b-t)}\right]_{r}~\label{eq.rtc.assume-reflect}
\ee
The last condition is:
\be
\left[e^{\lambda V(a,b)}\right]_{r}\text{ and }\left[O_{L}(a)e^{\lambda V(a,b)}\right]_{r}\text{ do not depend on the circumference of the wedge.}
\label{eq.rtc.assume-local}\ee
\end{subequations}
Therefore, the subtractions involved in renormalizing 
operators need to depend on the operators being renormalized only, and not on the 
size of the wedge state on which they are inserted.  
To achieve this last property, in  \cite{Kiermaier:2007vu}, renormalization took two steps: in the first,
the infinities were cancelled by subtracting the two-point function; in the second,
the finite part of the two-point function (which depends on the width of the 
wedge) was compensated for.  In contrast, we use the divergent part of
the two-point function alone to compensate for the divergence and then
study the impact of the finite part on the renormalization scheme, 
which means that in our approach, \eq{eq.rtc.assume-local} is automatically
satisfied.

In addition to these explicitly stated conditions, a
very natural condition of translation invariance was 
also implied in  \cite{Kiermaier:2007vu}.

At this point, it is relevant to ask what classes of 
operators we need to provide a renormalization scheme for.  Clearly,
we need to be able to renormalize exponentials and their
products.   This is done order by order, so operators such as 
$V(a,b)^{n}$ must be renormalizable.  Further,  the action of the
BRST operator $Q_B$ on $V(a,b)$ ($Q_BV(t) = \frac{\partial}{\partial t} (cV(t))$)
immediately implies that $O_L(a) = \lambda cV(a) + {\cal O}(\lambda^2)$
and $O_R(b) = \lambda cV(b) + {\cal O}(\lambda^2)$.  Thus,
we must be able to at least write down such operators as
$\left [V(a)e^{\lambda V(a,b)} \right ]_r$.  In fact, we will
see that this is sufficient: we need to renormalize
products of exponentials of integrated operators with
possible insertions of a single unintegrated $V$ on
either the left, or the right, or both.
These  operators also arise naturally when derivatives
are taken, for example: $\frac{\partial}{\partial a}\left [e^{\lambda V(a,b)} \right ]_r$.  

Once we have decided on the renormalization scheme
for $\left [e^{\lambda V(a,b)} \right ]_r$, derivative operators
such as $Q_B \left [e^{\lambda V(a,b)} \right ]_r$
and $\frac{\partial}{\partial a} \left [e^{\lambda V(a,b)} \right ]_r$ will be fixed.
The choice of renormalization scheme for such 
operators as $\left [V(a)e^{\lambda V(a,b)} \right ]_r$
can influence the explicit form of operators $O_{R,L}$ and
the existence of natural properties such as
\be
\label{linearity-1}
\frac{\partial}{\partial a}
\left [e^{\lambda V(a,b)} \right ]_r ~~ \overset{?}{=}~~ 
-\left [V(a)e^{\lambda V(a,b)} \right ]_r~,
\ee
but it does not change $Q_B \left [e^{\lambda V(a,b)} \right ]_r$
or $\frac{\partial}{\partial a} \left [e^{\lambda V(a,b)} \right ]_r$ themselves.
In other words, our choice of renormalization scheme
for operators with unintegrated insertions will not
affect the SFT solution.  However, it does affect
the linearity of the renormalization scheme (for example,
property \eq{linearity-1}).

We then need to ask: do the set of assumptions \eq{eq.rtc.assume-all}
imply that the renormalization scheme is linear?
The answer is that the replacement condition (\ref{eq.rtc.assume-reflect})
can be interpreted as a statement about linearity.  If we accept that
\be 
\ldots e^{\lambda V(a,c)}\ldots = 
\ldots e^{\lambda V(a,b)}e^{\lambda V(b,c)}\ldots
\ee
(a statement about the singular operators and not about 
the renormalization), then the replacement condition seems to
be a tautology.  Its true meaning is revealed when
we rewrite it order by order
\begin{subequations}
\be
\frac{1}{n!}\left[V(a,c)^{n}\right]_{r}=
\left[\sum_{j=0}^{n}\frac{1}{j!(n-j)!}V(a,b)^{j}V(b,c)^{n-j}\right]_{r}~
\ee
and then bring the combinatorial sum outside the renormalization:
\be\label{eq.rtc.as-replace-expanded}
\frac{1}{n!}\left[V(a,c)^{n}\right]_{r}=
\sum_{j=0}^{n}\frac{1}{j!(n-j)!}\left[V(a,b)^{j}V(b,c)^{n-j}\right]_{r}~.
\ee
\end{subequations}
Viewed this way, the replacement condition 
becomes a nontrivial statement about linearity
of the renormalization scheme when applied to
the exponentials and their products.  
We will see that this condition places
restrictions on possible renormalization schemes.

Repeated application of the replacement condition 
implies linearity for all operators of the form 
$V^{(n_1)}(a_1,b_1) \ldots V^{(n_k)}(a_k,b_k)$
with $a_1 < b_1 \leq a_2 < b_2 \leq a_3 \ldots \leq a_k < b_k$ as
long as the lengths $b_i-a_i$ of the intervals involved  are all 
finite.  It also applies to operators with an extra
insertion of an unintegrated operator on either the
left, the right or both (for example:
$V(a_0)V^{(n_1)}(a_1,b_1) \ldots V^{(n_k)}(a_k,b_k)$
where $a_0 \leq a_1$), but with one restriction:
in all the parts of the sum, the unintegrated
operator must be inserted \emph{at the same point}.
Therefore, the replacement condition does not imply
such linearity properties as
\eq{linearity-1}, which requires
that linearity be extended to addition of operators
where  the unintegrated
operator is inserted at different points.
Such `extended' linearity holds only for some
choices of renormalization schemes involving
unintegrated operator insertions.
Linearity beyond the replacement condition does not seem necessary
to construct the SFT solution, and does
not affect the details of this solution.
However, it is implicitly assumed in 
the analysis of conformal properties of the renormalized operator,
for example in \cite{Kiermaier:2007ki} (for more details, see section
\ref{small-operators-and-linearity}).

Our initial approach to renormalization will be to
consider products of integrated operators,
regulate them with a cut off by modifying the
domain of integration so that all insertions are separated
by a minimum distance of $\epsilon$, and introduce
counterterms that cancel the divergences when $\epsilon$
approaches zero.  We will modify the 
domain of  integration so that the insertions of integrated operators
are separated by at least $\epsilon$ from any fixed insertions.
We will use the notation $\left( ~\right)_{\epsilon}$ to denote
regulated operators, for example
\begin{subequations}\begin{align}
\left(V(a,b)^{2}\right)_{\epsilon}&=
\int_{a}^{b-\epsilon}dt_{1}\int_{t_{1}+\epsilon}^{b}dt_{2}~V(t_{1})V(t_{2})
~+~\int_{a+\epsilon}^{b}dt_{1}\int_{a}^{t_{1}-\epsilon}dt_{2}~V(t_{1})V(t_{2})~,
\label{eq.rtc.Vab2.epsilon}\\
\left(V(a,b)V(b,c)\right)_{\epsilon}&=\int_{a}^{b}dt_{1}\int_{\mathrm{max}(b,t_{1}+\epsilon)}^{c}dt_{2}~ V(t_{1})V(t_{2})
\label{eq.rtc.Vedge.epsilon}\\
\intertext{and}
\left(V(a)V(a,b)\right)_{\epsilon}&=
V(a)~\int_{a+\epsilon}^{b}dt~V(t)~, \label{eq.rtc.Vleft.epsilon}~.
\end{align}\end{subequations}

A crucial property of our regularization is that it is linear.
To show this, consider the most general operator with
$n$ $V$-insertions, $\int_M V(t_1) \ldots V(t_n)$ where $M$ is some measure on $\mathbb R^n$.
For example, $V(a)V(a,b)^2$ is associated with a uniform measure on $\{a\}\times(a,b)\times(a,b)$
(where $\{a\}$ is a point and $(a,b)$ is an interval.)
When adding two such operators, we simply add the corresponding measures.
The map $A \rightarrow (A)_\epsilon$ acts on the measure $M(A)$ associated with $A$ 
by setting it to zero for any point $(t_1, \ldots, t_n)$ such that $|t_i-t_j|<\epsilon$
and leaving it unchanged otherwise.  Denote this map by $U_\epsilon: M(A) \rightarrow M(A_\epsilon)$.
Since the action of the map $U_\epsilon$ on any given point within a measure depends only on
the coordinates of that point, we have that $U_\epsilon(M + \tilde M) =  U_\epsilon(M) +
U_\epsilon(\tilde M)$.  
Thus, $A \rightarrow (A)_\epsilon$ is a linear map for any operator.
This linearity property will be important to ensure that our renormalization
satisfies the replacement condition.

\begin{figure}
\centering
\subfloat[]{\includegraphics[width=0.4\textwidth]{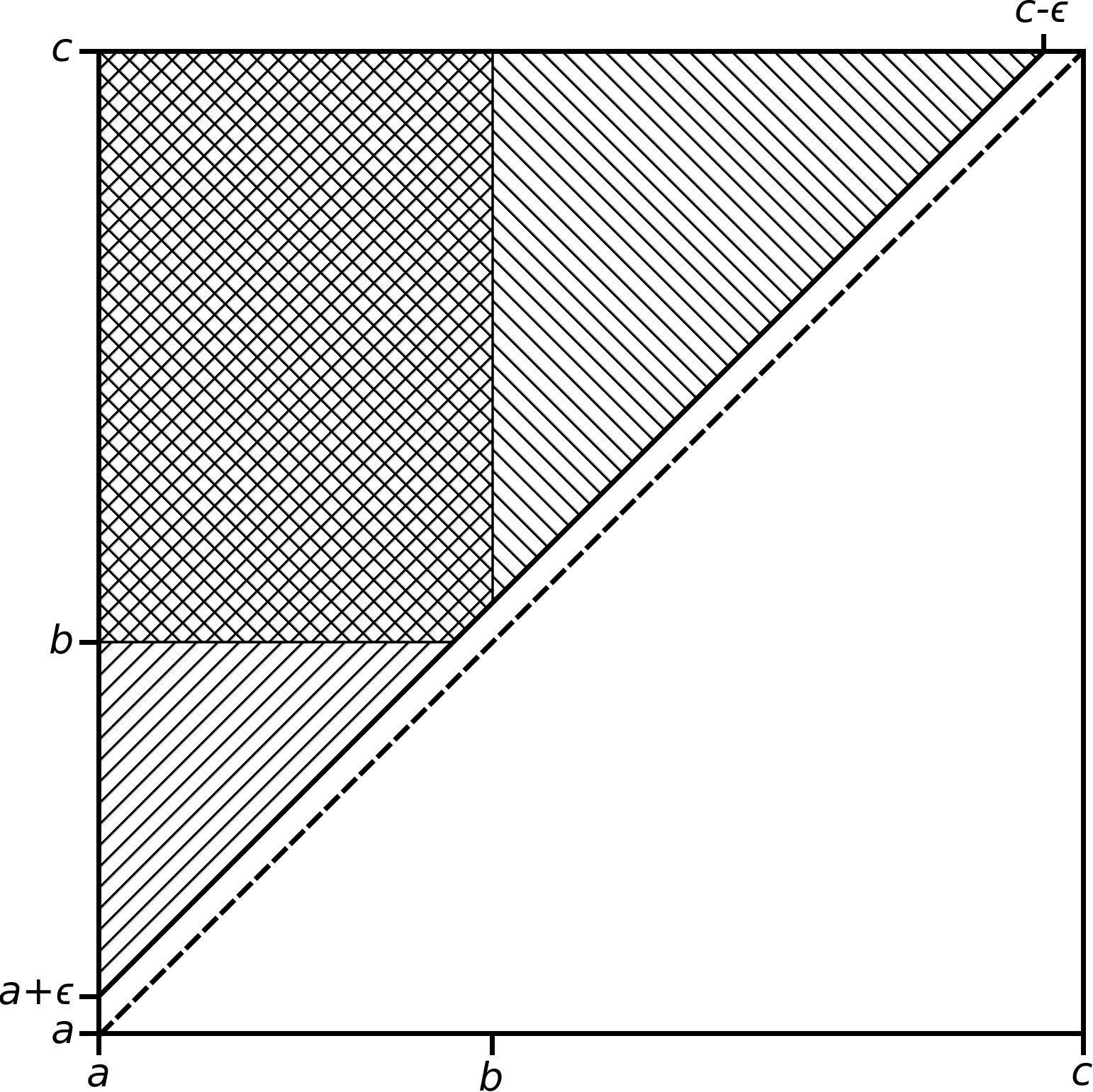}}
\hspace{0.1\textwidth}
\subfloat[]{\includegraphics[width=0.4\textwidth]{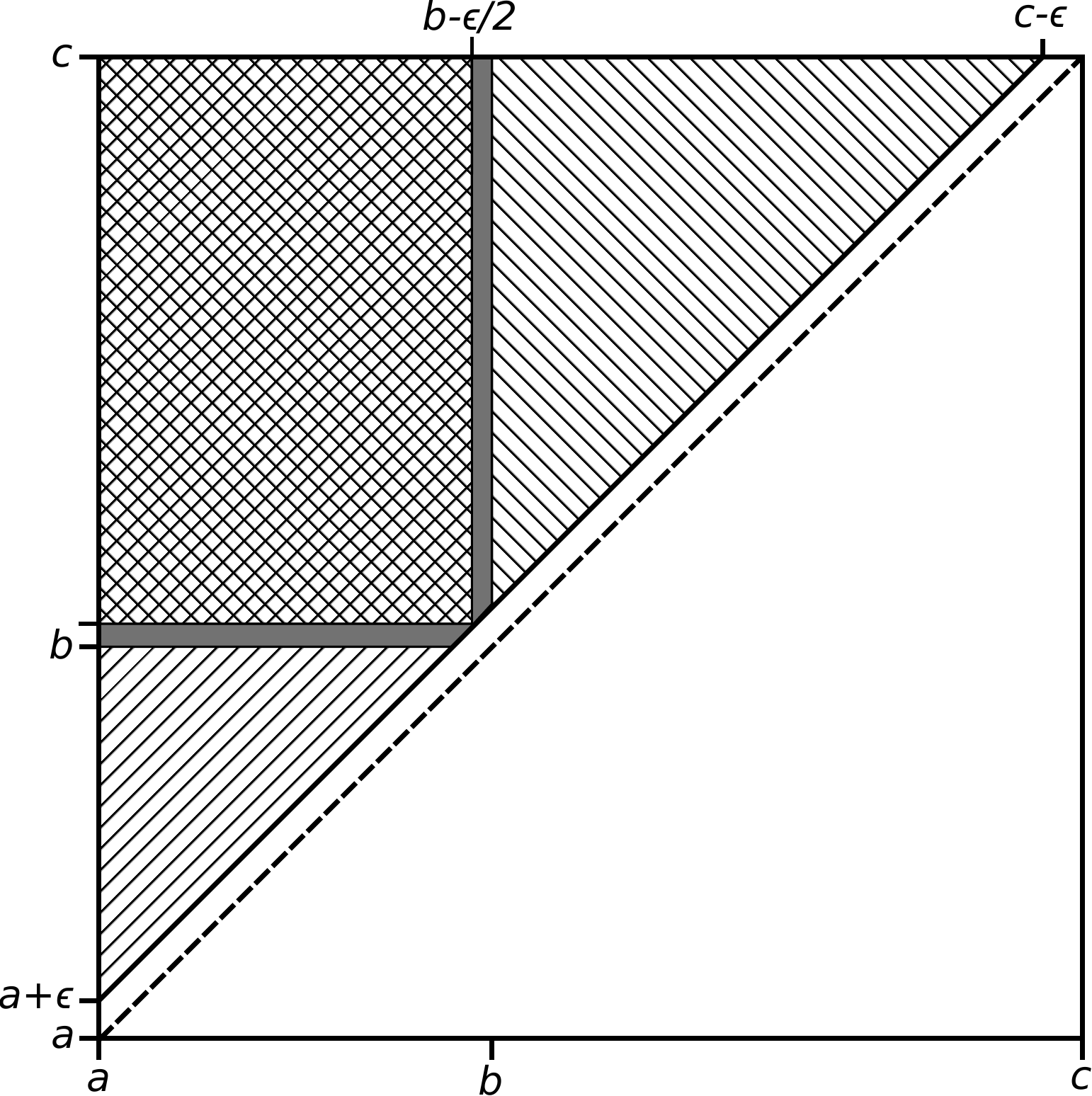}}
\caption[Comparison of two choices of integration region used for the 
renormalization of integrated operators]
{The regions of integration used in
$\left(V(a,b)^{2}\right)_{\epsilon}$ and 
$\left(V(b,c)^{2}\right)_{\epsilon}$ (diagonal hatching), and the 
region of $\left(V(a,b)V(b,c)\right)_{\epsilon}$ (cross-hatched):  (a) 
using our prescription for the renormalizations. (b) using the 
prescription of \cite{Kiermaier:2007vu}, equation \eq{edge-KO}.  The difference is the gray 
strips, which are not covered using the latter choice.  The dashed line 
indicates the location of a singularity due to colliding operators.}
\label{fig.rtc.int-regions}
\end{figure}

We should point out that our implementation here differs slightly from
\cite{Kiermaier:2007vu}.  In particular, the $\epsilon$-regularization of the operator $V(a,b)V(b,c)$ in
that work was not linear, being equivalent to
\be
\left(V(a,b)V(b,c)\right)_{\epsilon}^{\text{KO}}=\int_{a}^{b-\frac{\epsilon}{2}}dt_{1}\int_{b+\frac{\epsilon}{2}}^{c}dt_{2}V(t_{1})V(t_{2})~.
\label{edge-KO}
\ee
The difference between our definition \eq{eq.rtc.Vedge.epsilon} 
and the above equation is illustrated in figure \ref{fig.rtc.int-regions}.
This lack of linearity makes it difficult to see whether there
exists a complete renormalization scheme consistent with the assumption
\eq{eq.rtc.assume-replace} using the approach in \cite{Kiermaier:2007vu}.

In the $\epsilon \rightarrow 0$ limit, finite operators can be constructed by canceling the
divergences in the regulated operators with counterterms, so that generally
$[A]_r = (A)_\epsilon - \mathrm{counterterms}$.  While the divergent 
part of the counterterms is fixed by the OPEs of the operators in question,
the finite part is constrained only by the assumptions \eq{eq.rtc.assume-all}
and we will see that there is considerable freedom there.

\section{Renormalization of operators quadratic in $V$} 
\label{sec.rtc.quadratic}

In this section, we begin to construct a general
renormalization scheme, starting with the simplest
nontrivial situation: operators quadratic in $V$.
We will discuss some (but not all) of the conditions
\eq{eq.rtc.assume-all}.  Those we do not discuss in
this section will be proved in all generality
in section \ref{sec.rtc.all-orders}.

To start with, setting any other operator insertions aside, consider the operator
$V(a) V(a,b)$.  The corresponding $\epsilon$-regulated operator has the following
behaviour for small $\epsilon$:
\be \label{eq.rtc.left-OPE}
\left ( V(a) V(a,b) \right)_\epsilon = \frac{1}{\epsilon}
~+~\text{terms finite when $\epsilon$ goes to zero.}
\ee
Therefore, the corresponding renormalized operator can be 
defined as
\be \label{eq.rtc.left-ren-defn}
\left[V(a)V(a,b)\right]_{r}=\lim_{\epsilon\rightarrow 0}\left[
\left(V(a)V(a,b)\right)_{\epsilon} ~-~G^{L}_{ab}\right]
\ee
where the counterterm $G^{L}_{ab}$ is given by
\be
G^{L}_{ab}=\frac{1}{\epsilon}+C^{L}_{b-a}~. \label{eq.rtc.left-G-def}
\ee
Note that, by translation invariance, the counterterm $G^L_{ab}$
depends on $a$ and $b$ only through the difference $b-a$.
Discussion for $G^R_{ab}$ parallels that of $G^L_{ab}$.

Now, consider another operator requiring regularization,
$V(a,b)^2$.  We have:
\be
\half \left (V(a,b)^2 \right )_\epsilon=\frac{b-a}{\epsilon}+\ln\epsilon
~+~\text{terms finite when $\epsilon$ goes to zero.}
\ee
(Our factor of $\half$ is convenient when evaluating the integral:
the two components in definition \eq{eq.rtc.Vab2.epsilon} are equal to each other
and therefore either one of them is equal to $\half \left (V(a,b)^2 \right )_\epsilon$.)
The corresponding renormalized operator is then given by
\be \label{eq.rtc.double-ren-defn}
 \left[V(a,b)^{2}\right]_{r}=\lim_{\epsilon\rightarrow0}\left[\left(V(a,b)^{2}\right)_{\epsilon}~-~2G^{D}_{ab}\right]~,
\ee
where the ``double'' counterterm $G^{D}_{ab}$ for doubly integrated operators is 
\be
G^{D}_{ab}=\frac{b-a}{\epsilon}+\ln\epsilon+C^{D}_{b-a}~.\label{eq.rtc.double-G-def}
\ee
By a similar process, we also define
\be \label{eq.rtc.abc-ren-defn}
\left[V(a,b)V(b,c)\right]_{r}=\lim_{\epsilon\rightarrow0}\left[ \left(V(a,b)V(b,c)\right)_{\epsilon}
~-~G^{E}_{abc} \right ]~,
\ee
where $G^{E}_{abc}$ is the ``edge'' counterterm for operators meeting only at 
a single shared edge,
\be
G^{E}_{abc} = -\ln\epsilon+C^{E}_{c-b,b-a}~.\label{eq.rtc.edge-G-def}
\ee

As we have discussed already in section \ref{sec.rtc.ko},
our choice of $C^{L,R}_{ab}$ cannot influence the SFT solution,
but the choice of $C^{D}_{b-a}$ and $C^{E}_{c-b,b-a}$ certainly can.
We now find restrictions on these finite parts of the 
counterterms due to the replacement condition 
{\eq{eq.rtc.assume-replace} and the
more general assumption of linearity.

\subsection{Replacement condition  \eq{eq.rtc.assume-replace}} 
\label{sec.rtc.assume-cd}

We begin by using condition (\ref{eq.rtc.assume-replace})
with an insertion of $V(a)$ on the left:
\be
\left[V(a) e^{\lambda V(a,c)}\right]_{r}
=\left[V(a) e^{\lambda V(a,b)}e^{\lambda V(b,c)}\right]_{r}~,
\ee 
which to first order in $\lambda$  implies that
\begin{subequations}\begin{gather}
\left[V(a)V(a,c)\right]_{r}=\left[V(a)V(a,b)\right]_{r}+\left[V(a)V(b,c)\right]_{r}
= \left[V(a)V(a,b)\right]_{r}+V(a) V(b,c)~.\\
\intertext{For the last term we have used the factorization condition 
\eq{eq.rtc.assume-factor} to remove the renormalization.  Since it is trivially true that}
\left(V(a)V(a,c)\right)_{\epsilon}= \left(V(a)V(a,b)\right)_{\epsilon}
+V(a) V(b,c)~,\\
\intertext{we obtain that} 
C^{L}_{c-a}=C^{L}_{b-a}=C^L~.
\end{gather}\end{subequations}
The finite part of the counterterm for $\left[V(a)V(a,b)\right]_{r}$
is a constant $C^L$ which does not depend on the size of the integration region.  

A similar argument, condition (\ref{eq.rtc.assume-replace})
with an integrated operator inserted on the left:
\be
\left[e^{\lambda_1(a,b)} e^{\lambda V(b,d)}\right]_{r}
=\left[e^{\lambda_1(a,b)} e^{\lambda V(b,c)}e^{\lambda V(c,d)}\right]_{r}~,
\ee 
when expanded to first order in $\lambda$ and $\lambda_1$, gives
\begin{subequations}\begin{gather}
\left[V(a,b)V(b,d)\right]_{r}
= \left[V(a,b)V(b,c)\right]_{r}+V(a,b)V(c,d)~,\\
\intertext{which, together with}
\left(V(a,b)V(b,d)\right)_{\epsilon}= \left(V(a,b)V(b,c)\right)_{\epsilon}
+V(a,b) V(c,d)~,\\
\intertext{yields} 
C^{E}_{d-b,b-a}=C^{E}_{c-b,b-a}~.
\end{gather}\end{subequations}
This, together with a similar statement given when the extra
integrated operator is inserted on the right, implies that $C^{E}_{c-b,b-a}=C^{E}$ is
independent of the values of $a$, $b$ and $c$.

Next, we examine  condition (\ref{eq.rtc.assume-replace}) without any extra
insertions at second order in $\lambda$:
\begin{subequations}
\be
\left[V(a,c)^{2}\right]_{r}=\left[V(a,b)^{2}\right]_{r}+\left[V(b,c)^{2}\right]_{r}+2\left[V(a,b)V(b,c)\right]_{r}.
\ee
Since, as follows trivially from the 
linearity of $(\ \cdot\ )_{\epsilon}$, 
\be
\left(V(a,c)^{2}\right)_{\epsilon}=\left(V(a,b)^{2}\right)_{\epsilon}+\left(V(b,c)^{2}\right)_{\epsilon}+2\left(V(a,b)V(b,c)\right)_{\epsilon}~,
\ee
after canceling the $\epsilon$-dependent terms, we are left with
\be
C^{D}_{c-a}-C^{D}_{c-b}-C^{D}_{b-a}-C^{E}=0 \label{eq.rtc.const-condition}~.
\ee
\end{subequations}
Therefore the finite counterterm must be of the form
$C^{D}_{\Delta t}=C_{0}+C_{1}\Delta t$, and we must have
$C^{E}=-C_{0}$.  

At this point, our renormalization scheme is parametrized by
4 parameters: $C_0$, $C_1$, $C^R$ and $C^L$.  We will see that
$C^R$ and $C^L$ do not affect the eventual SFT solution (and
we could have set them to zero without a loss of generality) but
that $C_0$ and $C_1$ do.  It is worth mentioning that, at this order, our
scheme reproduces that of \cite{Kiermaier:2007vu} if we take
$C_1=0$ and $C_0=-1$, though comparison for operators with multiple integration
regions is complicated by
the discrepancy described in section \ref{sec.rtc.ko}.

\subsection{The BRST assumptions \eq{eq.rtc.assume-Q1} and \eq{eq.rtc.assume-Q2}}}

These two assumptions are easily proven at second order in $\lambda$.
Throughout this section we will 
omit  the limit $\epsilon\rightarrow 0$, and it should be 
inferred. 
Recall that the first BRST assumption is
\be
Q_{B}\left[e^{\lambda V(a,b)}\right]_{r}=\left[e^{\lambda V(a,b)}O_{R}(b)\right]_{r}-\left[O_{L}(a)e^{\lambda V(a,b)}\right]_{r}\tag{\ref{eq.rtc.assume-Q1}}~.
\ee
At second order (in $\lambda$) this statement reads as 
\be \label{eq.rtc.Q1-2ndorder-statement}
\frac{1}{2}Q_{B}\left[V(a,b)^{2}\right]_{r}=\sum_{n=0}^{2}\frac{1}{(2-n)!}\left(\left[V(a,b)^{2-n}O_{R}^{(n)}(b)\right]_{r}-\left[O_{L}^{(n)}(a)V(a,b)^{(2-n)}\right]_{r}\right)~,
\ee
where $O_{L/R}=\sum_{n}\lambda^{n}O_{L/R}^{(n)}$ is a local operator 
to be determined.  The behaviour of primitive 
operators when acted on by the BRST charge is not difficult to 
determine by integrating the BRST current on a contour about the 
operator in question.  The results we will need are 
\be\begin{split}
Q_{B}V(t)=\partial_{t}(cV(t))~,\qquad
Q_{B}(cV(t))=0~,\\
Q_{B}c(t)=c\partial c(t)~,\qquad
Q_{B}\partial c(t)=c\partial^{2}c(t)~.
\end{split}\label{eq.rtc.Q-primitives}\ee
Using these and the definition of the renormalization scheme, we can 
start working out the left hand side of 
\eq{eq.rtc.Q1-2ndorder-statement} explicitly.
\begin{subequations}\begin{align}
\frac{1}{2}Q_{B}\left[V(a,b)^{2}\right]_{r}&=\frac{1}{2}Q_{B}\left(V(a,b)^{2}\right)_{\epsilon}\\
&=Q_{B}\int_{a}^{b-\epsilon}dt_{1}\int_{t_{1}+\epsilon}^{b}dt_{2}V(t_{1})V(t_{2})\\
&=\int_{a}^{b-\epsilon}dt_{1}\int_{t_{1}+\epsilon}^{b}dt_{2}\left(\partial_{t_{1}}cV(t_{1})V(t_{2})+V(t_{1})\partial_{t_{2}}cV(t_{2})\right) \label{eq.rtc.Q1-2ndorder-start}
\end{align}\end{subequations}

The next step is to integrate by parts: 
\begin{subequations}\begin{align}
\frac{1}{2}Q_{B}&\left[V(a,b)^{2}\right]_{r}
=\int_{a}^{b-\epsilon}dt~V(t)\left(cV(b)-cV(t+\epsilon)\right)+\int_{a+\epsilon}^{b}dt\left(cV(t-\epsilon)-cV(a)\right)V(t)\\
&=V(a,b-\epsilon)cV(b)-cV(a)V(a+\epsilon,b)+\int_{a+\epsilon}^{b}dt\left(cV(t-\epsilon)V(t)-V(t-\epsilon)cV(t)\right)
\end{align}\end{subequations}

The integral in the last term is equal to (neglecting
all terms that go to zero as $\epsilon \rightarrow 0$),
\begin{subequations}\begin{align}
\int_{a+\epsilon}^{b}dt~V(t-\epsilon)V(t)\left(c(t-\epsilon)-c(t)\right)
&=\int_{a+\epsilon}^{b}dt~\frac{1}{\epsilon^{2}}\left(-\epsilon\partial c(t)+\frac{\epsilon^{2}}{2}\partial^{2}c(t)\right)\\
&=-\frac{1}{\epsilon}\left(c(b)-c(a+\epsilon)\right)+\frac{1}{2}\left(\partial c(b)-\partial c(a)\right)\\
&=-\frac{1}{\epsilon}\left(c(b)-c(a)\right) +\frac{1}{2}\left(\partial c(b)+\partial c(a)\right)~.
\end{align}\end{subequations}
Thus:
\begin{subequations}\begin{align}
\frac{1}{2}Q_{B}\left[V(a,b)^{2}\right]_{r}
&=V(a,b-\epsilon)cV(b)-\frac{c(b)}{\epsilon}+\frac{1}{2}\partial c(b)
-cV(a)V(a+\epsilon,b)+\frac{c(a)}{\epsilon}+\frac{1}{2}\partial c(a)\\
&=\left[V(a,b)cV(b)\right]_{r}+C^{R}c(b)+\frac{1}{2}\partial c(b)
-\left[cV(a)V(a,b)\right]_{r}-C^{L}c(a)+\frac{1}{2}\partial c(a)
\end{align}\end{subequations}
This has the form of \eq{eq.rtc.Q1-2ndorder-statement} where 
\be
O_{R}(b)=\lambda cV(b)+\frac{\lambda^{2}}{2}\partial c(b)+\lambda^{2}C^{R}c(b)~,\quad
O_{L}(a)=\lambda cV(a)-\frac{\lambda^{2}}{2}\partial c(a)+\lambda^{2}C^{L}c(a)~.
\label{OLR}
\ee
Our operators $O_{R,L}$ depend explicitly on the renormalization parameters $C^{L,R}$.
However, this dependence only serves to cancel the dependence of the renormalization
scheme on these parameters, so that in fact $\left [O_L(a) e^{\lambda V(a,b)}\right]_r$ 
and $\left [e^{\lambda V(a,b)}O_R(b) \right]_r$  are independent of $C^{L,R}$ (as they must be).
Notice that condition \eq{eq.rtc.assume-reflect} implies that
\be
\left [O_{L}(a)e^{\lambda V(a,b)}\right]_{r} ~ \overset{t\rightarrow (a+b)-t}{\rightarrow}  ~~
\left [e^{\lambda V(a,b)}O_{R}(b)\right]_{r}~,
\label{reflection-assumption-with-o}
\ee
but this does not restrict $C^R$ and $C^L$ to be equal.

For the second BRST condition, \eq{eq.rtc.assume-Q2}, at this order
we just need to show that
\be\label{eq.rtc.Q2-quadratic-goal}
Q_{B}\left(\left[cV(a)V(a,b)\right]_{r}-\frac{1}{2}\partial c(a)+C^{L}c(a) \right )
=-cV(a)cV(b)~.
\ee
This is shown as follows:
\begin{multline}
Q_{B}\left(\left[cV(a)V(a,b)\right]_{r}-\frac{1}{2}\partial c(a)+C^{L}c(a) \right ) \\
= Q_B\left(cV(a)\int_{a+\epsilon}^b \! dt V(t) - c(a)G^L_{ab} -\frac{1}{2}\partial c(a)+C^Lc(a)\right ) \\
= -cV(a)\left (cV(b) - cV(a+\epsilon)\right ) - \frac{c\partial c(a)}{\epsilon} - C^Lc \partial c(a)
- \half c\partial^2 c(a) + C^L c\partial c(a)\\
\end{multline}
Writing $V(a)V(a+\epsilon) = \epsilon^{-2}$+finite terms and $c(a)c(a+\epsilon) = \epsilon c\partial c(a) + \half \epsilon^2 c\partial^2 c$, several terms cancel and
we end up with
\be
Q_{B}\left(\left[cV(a)V(a,b)\right]_{r}-\frac{1}{2}\partial c(a)+C^{L}c(a) \right )
=-cV(a)cV(b)
\ee
Thus, the second BRST condition is satisfied at this order as well.

\subsection{Linearity and boundary condition changing operators}
\label{small-operators-and-linearity}

In this section, we discuss linearity of the renormalization
scheme beyond the replacement condition.  We investigate the
consequences of such natural and related\footnote{ 
By the fundamental theorem of calculus.}
assumptions as
\begin{subequations}
\be
\int_a^b dt [V(t) V(t,c)]_r~\overset{?}{=}~ \half [V(a,b)^2]_r + [V(a,b)V(b,c)]_r
\label{linearity-integral}
\ee
and
\be
\frac{\partial}{\partial a}[V(a,b)^2]_r ~\overset{?}{=}~ -2 \left [V(a) V(a,b) \right]_r~.
\label{linearity-derivative}
\ee
\end{subequations}
One can ask why we would be interested in such conditions,
given that they don't seem to be needed to construct a
SFT solution.  The answer is that these properties are 
related to the conformal properties of the corresponding
boundary condition changing (bcc) operator. 

We might assume (as has been the focus of recent work, for example 
\cite{Kiermaier:2010cf,Maccaferri:2014cpa,Erler:2014eqa}),
that the point where the boundary condition is changed behaves
as if a bcc operator $\sigma$ was inserted there.  If this operator
is primary and has conformal weight $h(\lambda)$, we would expect that \cite{Kiermaier:2007ki}
\be
\left [O_L(a) e^{\lambda V(a,b)} \right ]_r = -c(a) \frac{\partial}{\partial a} \left [e^{\lambda V(a,b)} \right ]_r
- h(\lambda) \partial c(a)\left [e^{\lambda V(a,b)} \right ]_r~.
\label{conformal}
\ee
Thus, to compare with our formula for $O_L$, we need to know the form
that the derivative takes.  However, as the equivalent assumption \eq{linearity-integral} is easier to investigate,
we start there.

The definition of the $\epsilon$-regularization implies that
\be
\int_a^b dt (V(t) V(t,c))_\epsilon= \half (V(a,b)^2)_\epsilon + 
(V(a,b)V(b,c))_\epsilon~.
\ee
Subtracting this expression from equation (\ref{linearity-integral}), we obtain
\be
C^L(b-a) = C_0 + C_1(b-a) + C^{E}~.
\ee
Since this equation should be true for arbitrary $a$ and $b$, we obtain
a new constraint $C^L = C_1$ and confirm our previous result that
$C_0 = -C^{E}$.  A similar argument with the unintegrated
operator on the right implies that $C^R = C_1$.  

Let us now examine the statement about a derivative, \eq{linearity-derivative}. 
 We look at
\begin{subequations}
\begin{align}
\partial_a[V(a,b)^2]_r &= \lim_{\Delta \rightarrow 0} \frac{[V(a,b)^2]_r - [V(a-\Delta,b)^2]_r}{\Delta}\\  & = 
-\lim_{\Delta \rightarrow 0} \frac{[V(a-\Delta,a)^2]_r + 2[V(a-\Delta,a)V(a,b)]_r}{\Delta} 
\\ & =
-\lim_{\Delta \rightarrow 0}\lim_{\epsilon\rightarrow0} \frac{\left(V(a-\Delta,a)^2\right)_\epsilon + 
2\left(V(a-\Delta,a)V(a,b)\right)_\epsilon - 2\frac{\Delta}{\epsilon} - 2C_1\Delta}{\Delta} ~,
\end{align}
where we have used that $C_0=-C^E$.  Now, we carefully examine the integration
regions for the two $\epsilon$-regulated expressions and discover that they can
be recombined to give
\begin{align}
&-2\lim_{\Delta \rightarrow 0}\lim_{\epsilon\rightarrow0} \frac{\int_0^\Delta dz V(a-z) \int_{a-z+\epsilon}^b dt V(t)
 - \frac{\Delta}{\epsilon} - C_1\Delta}{\Delta} 
\\ & = -2\lim_{\Delta \rightarrow 0}\lim_{\epsilon\rightarrow0} \frac{\int_0^\Delta dz \left \{ \left(V(a-z)V(a-z,b)\right)_\epsilon
- \frac{1}{\epsilon} - C^L \right \} }{\Delta} ~+~2(C_1-C^L)
\\ & = -2\lim_{\Delta \rightarrow 0} \frac{\int_0^\Delta dz  \left[V(a-z)V(a-z,b)\right]_r
}{\Delta} ~+~2(C_1-C^L)
\\ & = -2  \left[V(a)V(a,b)\right]_r ~+~2(C_1-C^L)~.
\end{align}
\end{subequations}
In the last line we have used the fact that $\left[V(a-z)V(a-z,b)\right]_r$
is a finite and smooth function of $z$ to write
\be
\int _0^\Delta dz \left [V(a-z) V(a-z,b) \right ]_r = \Delta \left [V(a) V(a,b) \right]_r
+ \mathcal{O}(\Delta^2)~.
\ee
Thus, we obtained a formula for the derivative:
\be
\frac{\partial}{\partial a}[V(a,b)^2]_r = -2  \left[V(a)V(a,b)\right]_r ~+~2(C_1-C^L)~.
\label{derivative-answer}
\ee  
The linear result \eq{linearity-derivative} holds
only for $C_1 = C^L$, which is (unsurprisingly) the same condition that we obtained from
requiring \eq{linearity-integral}.  

Now, using equation \eq{derivative-answer}, we can compare equations \eq{conformal}
and \eq{OLR}.  We see that the bcc operator must have
conformal weight $\half \lambda^2$ (this was already discussed in \cite{Kiermaier:2007ki})
and that $C_1$ must be zero.  Thus, interestingly, while we can 
assume any $C_1$ to construct a SFT, only for $C_1=0$ will this solution
have a \emph{primary} bcc operator.

In our computation of the derivative, we were careful to not 
bring the $\Delta \rightarrow 0$ limit inside the regularization bracket $[\ldots]_r$,
as pathologies can develop when doing so.  For example,
an explicit computation using the $VV$ OPE gives that
\be
\label{eq.rtc.smallintop}
[V(a-\Delta,a)^2]_r = -2 (\ln \Delta + 1 + C_0 + C_1\Delta) + \mathcal{O}(\Delta^2)~,
\ee
which is infinite in the $\Delta \rightarrow 0$ limit.
Naively, $\lim_{\Delta \rightarrow 0} V(a-\Delta,a)^2$ might be thought 
to be zero, since the operators are integrated over a set whose measure approaches zero.
But this too is suspect, as it's not clear what $ \lim_{\Delta \rightarrow 0}V(a-\Delta,a)^2$ means 
without any regulation.  Further, at fixed $\epsilon$,
 $\lim_{\Delta \rightarrow 0} \left(V(a-\Delta,a)^2\right)_\epsilon = 0$,
so we could write that
\be
\left [\lim_{\Delta \rightarrow 0} V(a-\Delta,a)^2 \right ]_r = 
- \lim_{\epsilon \rightarrow 0 } ~\lim_{\Delta \rightarrow 0}~2G^D_{\Delta} = 
- \lim_{\epsilon \rightarrow 0 } \left ( 2\ln \epsilon + 2C_0\right )~,
\ee
which is again infinite.

The divergence for $\Delta \rightarrow 0$ 
in equation \eq{eq.rtc.smallintop} is
necessary and it has a simple interpretation in terms of the OPE of the 
corresponding boundary condition changing (bcc) operator, 
\be
\bar \sigma(s) \sigma(0) ~\sim~\frac{1}{s^{2h}}~+~\ldots~,
\ee
where $h$ is the conformal weight of the bcc operator $\sigma$.
As we already saw, the conformal weight is
related to $\lambda$ by $2h=\lambda^2$.  
At the lowest nontrivial order in $\lambda$, the divergent
part of the above OPE is
\be
\bar \sigma(s) \sigma(0) 
= e^{-\lambda^2 \ln s} ~+~ \ldots = -\lambda^2 \ln s  + 
\mathrm{terms~that~are~finite~or~higher~order~in~}\lambda~.
\ee
The term $-\lambda^2 \ln s$ is exactly what we obtained in
equation \eq{eq.rtc.smallintop}:
\be
\bar \sigma(s) \sigma(0) = \left [ e^{\lambda V(0,s)} \right ]_r
= \half \lambda^2  \left [ {V(0,s)^2} \right ]_r  + \ldots
= -\lambda^2 \ln s  + \ldots
\ee

Finally, we conclude this subsection with a warning.
Naively, the following two regulated operators should be equal:
\be
\left [ V(a,b)^2 \right ]_r ~~\mathrm{and}~~
\int_a^b dt~\left ( 
\left [V(t) V(t,b) \right ]_r + \left [ V(a,t) V(t) \right ]_r 
\right )~,
\label{compare-the-two}
\ee
However, it is easy to see that 
\be
\int_a^b dt~\left ( 
\left [V(t) V(t,b) \right ]_r + \left [ V(a,t) V(t) \right ]_r 
\right ) = \left ( V(a,b)^2 \right )_\epsilon - \frac{b-a}{\epsilon} -
(C^R+C^L)(b-a)~,
\label{integrated-edge-infinite}
\ee
which is not the same as $\left [ V(a,b)^2 \right ]_r $.
In particular, \eq{integrated-edge-infinite} is missing the 
divergent $\ln \epsilon$ part of the counterterm,
so it's not even finite.  What went wrong?  On the LHS of equation
\eq{integrated-edge-infinite} we included a small operator
$\lim_{t\rightarrow b} V(t)V(t,b)$, which is divergent even
when regulated.  The two operators in \eq{compare-the-two}
would only be equal if we were able to commute the order of
integration and regularization, which fails when the 
operators involved are small. Notice that we were careful not 
to use small operators when we wrote down 
equation \eq{linearity-integral}.
~\\ \\
To summarize section \ref{sec.rtc.quadratic}, 
we have found that the factorization and replacement conditions
restrict possible renormalization schemes for two operators 
to
\begin{subequations}\begin{align}
G^L_{ab} &= \frac{1}{\epsilon} + C^L~,\quad G^R_{ab} = \frac{1}{\epsilon} + C^R~, \\
G^{D}_{ab} &= \frac{b-a}{\epsilon} + \ln \epsilon + C_0 + C_1(b-a)~, \\
G^{E}_{abc} &= -\ln \epsilon - C_0.
\end{align}\end{subequations}
Parameters $C^R$ and $C^L$ do not change the SFT solution
and could be set to zero without loss of generality.
Insisting on linearity implies a further condition that $C^L=C_1=C^R$,
while the bcc operator corresponding to the renormalized boundary 
deformation is primary only if $C_1 = 0$.  


\section{Third order}
\label{sec.rtc.cubic}

Before we plunge into a computation at all orders, we will
consider our renormalization scheme at third order, i.e.
the renormalization of a product of three $V$s.  

At this order, we define a regularized operator involving a single
integrated operator by following the same regularization pattern as we did for 
the quadratic operator:
\be
\left [V(a,b)^3\right]_r = 
\left(V(a,b)^3\right)_\epsilon~-~ 6G_{ab}^{(3),D}~ V(a,b)~,
\ee
where
\be
G_{ab}^{(3),D} = \frac{b-a}{\epsilon} +\ln\epsilon  + C_{b,a}^{(3),D}~.
\ee
The extra superscript $(3)$ indicates that these are the 
counterterms at third order.
We also define a regularized operator involving two integrated
operators:
\begin{subequations}
\be
\left [V(a,b)^2 V(b,c)\right]_r = 
\left(V(a,b)^2 V(b,c)\right)_\epsilon~-~ 2 G_{abc}^{(3), DE} V(b,c)~-
~2 G_{abc}^{(3),E} V(a,b)~,
\ee
and three operators:
\be
\left [V(a,b) V(b,c) V(c,d)\right]_r = 
\left(V(a,b) V(b,c) V(c,d)\right)_\epsilon~-~  G_{abcd}^{(3),EE} V(a,b)~-
G_{abcd}^{(3),EE} V(c,d)
\ee
\end{subequations}
where
\begin{subequations}\begin{align}
G_{abc}^{(3),E} &= -\ln\epsilon  + C_{abc}^{(3),E}~, \\
G_{abc}^{(3),DE} &= \frac{b-a}{\epsilon} +\ln\epsilon  + C_{abc}^{(3),DE}~, \\
G_{abcd}^{(3),EE} &= -\ln\epsilon  + C_{abcd}^{(3),EE}~.
\end{align}\end{subequations}
Notice that we have four new and potentially different counterterms.
Using translation invariance together with the factorization
and replacement conditions in a way similar to that presented in the quadratic case,
we can show that 
\begin{subequations}\begin{align}
C_{abc}^{(3),E} &=   -C_0~,  \\
C_{abce}^{(3),EE} &=  -C_0~,  \\
C_{ab}^{(3),D} &= (b-a)C_1 + C_0^{(3)}~,\\ 
C_{abc}^{(3),DE} &= (b-a)C_1 + C_0~,
\end{align}\end{subequations}
where the constants $C_1$ and $C_0$ are necessarily the
same as the ones used at quadratic order, but $C_0^{(3)}$
is a new independent constant.
One can check, by examining all combinations, that the replacement condition at
third order is satisfied for any value of $C_0^{(3)}$.

We also need to define renormalized operators involving 
unintegrated insertions; using factorization and replacement
conditions, these can be constrained to
\begin{subequations}\begin{align}
[V(a)V(a,b)^2]_r &= (V(a)V(a,b)^2)_\epsilon - 2V(a)G^{(3),DL}_{ab} - 2V(a,b) G^{(3),L}_{ab} ~,\\
[V(a)V(a,b)V(b)]_r &= (V(a)V(a,b)V(b))_\epsilon - V(a) G^{(3),RL}_{ab} -  V(b) G^{(3),LR}_{ab}~,\\
[V(a)V(a,b)V(b,c)]_r &= (V(a)V(a,b)V(b,c))_\epsilon - V(a) G^{(3),EL}_{abc} - V(b,c) G^{(3),LE}_{abc}~,
\end{align}\end{subequations}
where
\begin{subequations}\begin{align}
G^{(3),DL}_{ab}  &= \frac{b-a}{\epsilon} + \ln \epsilon + C_0^{(3),DL} + (b-a)C_1~, \\
G^{(3),L}_{ab}   &= \frac{1}{\epsilon} +  C^L~,\\
G^{(3),RL}_{ab}  &=  \frac{1}{\epsilon} +  C^R~,\\
G^{(3),LR}_{ab}  &=  \frac{1}{\epsilon} +  C^L~,\\
G^{(3),EL}_{abc} &=  -\ln \epsilon - C_0~,\\
G^{(3),LE}_{abc} &= \frac{1}{\epsilon} +  C^L~.
\end{align}\end{subequations}
There are two new constants: $C_0^{(3),DL}$ and its partner,
$C_0^{(3),DR}$.  Just like $C^L$ and $C^R$, however,
these constants cannot change the SFT solution and can only affect the form of
the BRST insertions $O^L$ and $O^R$.  For example,
an explicit computation shows that the first BRST condition
holds with corrected boundary operators:
\begin{subequations}\begin{align}
O_{L}(a)&=\lambda cV(a)-\frac{1}{2}\lambda^{2}\partial c(a)
+\lambda^{2}C^{L}c(a)+\lambda^{3}\left(C_{0}^{(3),DL}-C_{0}^{(3)}\right)cV(a)\\
\intertext{and}
O_{R}(b)&=\lambda cV(b)+\frac{1}{2}\lambda^{2}\partial c(b)
+\lambda^{2}C^{R}c(b)+\lambda^{3}\left(C_{0}^{(3),DR}-C_{0}^{(3)}\right)cV(b)~.
\end{align}\end{subequations}
As we did at quadratic order, this should be compared with
\be
\frac{\partial}{\partial {a}}
\left[V(a,b)^{3}\right]_{r}=-3\left[V(a)V(a,b)^{2}\right]_{r}+6V(a,b)\left(C_{1}-C^{L}\right)+6V(a)\left(C_{0}^{(3)}-C_{0}^{(3),DL}\right)~.
\ee
We see that the bcc operator corresponding to our solution is
still primary at this order as long as $C_1=0$.

At this order we did find one new free parameter that
can affect the SFT solution: $C_0^{(3)}$.  It is clear
that if we were to continue our order-by-order approach
to renormalization, we would find new free parameters.
However, at quartic and higher orders, this approach
in unwieldy: it is hard to write down the most general
renormalized operator that is demonstratively finite.
To study renormalization to all orders, we will no
longer try to study the space of all renormalizations and instead
focus on a particular renormalization scheme.  The
scheme we chose will have $C_0$ and $C_1$ as free
parameters, however we will not add new constants
at every order.  We will return to the question of
classifying all renormalization schemes in section
\ref{sec.discussion}.

\section{Renormalization to all orders}
\label{sec.rtc.all-orders}

In this section, we present an example renormalization scheme
at all orders.  Our scheme is demonstratively finite, and 
we prove that it satisfies all the  conditions set out in section
\ref{sec.rtc.ko}.  At second order, our scheme matches
that described in section \ref{sec.rtc.quadratic}, and
it has the same two free parameters, $C_0$ and $C_1$.

To define the full renormalization scheme, we need to 
consider what kind of singularities can appear when
considering products of three or more operators.  One
class of singularities appears when any two of these operators
are inserted at the same point; we can deal with this
class of singularities by recursively subtracting 
divergences that occur when any two operators are
inserted at the same point.
However, it is also possible to have additional singularities.  Since
the finite part of the OPE of any two operators that are close together
will contain operators other than the identity, another
operator inserted close can then have a singular OPE with these
operators.  In other words, we can have additional divergences caused by
three or more operators inserted at the same point.
Following  equation (4.10) of \cite{Kiermaier:2007vu}, we require that
such singularities are not present and restrict
our arguments to a class of operators $V$ such that
\be \label{eq.rtc.finiteness-from-OPE}
\exp\left(-\frac{1}{2}\int dt_{1}dt_{2}~\frac{1}{(t_{1}-t_{2})^{2}} 
\frac{\delta}{\delta V(t_{1})}\frac{\delta}{\delta V(t_{2})}\right)
\prod_{i}V(t_{i})
\ee
remains finite even when more than two of the coordinates $t_{i}$ 
collide simultaneously.  This implies that to decide whether
any renormalization scheme leads to a finite operator, we only
have to ensure that the renormalized operator stays finite
in the limit $t_i \rightarrow t_j$ for any pair of coordinates
$t_i$ and $t_j$. 
With this restriction in place, it is sufficient, for
composite operators with more than two factors, 
to subtract the divergence which results from any
two operators coming together.

Now, consider for example $\left(e^{\lambda V(a,b)}\right)_{\epsilon}$.  As
$\epsilon \rightarrow 0$, this operator diverges.  To regulate it, we might
propose an expression such as
\be \label{eq.rtc.highorder-one-factor}
\lim_{\epsilon\rightarrow0}~ 
e^{-\lambda^{2} G^{D}_{a,b}}~\left(e^{\lambda V(a,b)}\right)_{\epsilon}~.
\ee
Using the quadratic counterterm $G_{ab}^D$ to define the renormalization at all
orders would correspond to making many choices about 
finite parts of higher order counterterms, for example choosing $C_0^{(3)} = C_0$
at third order.  However, as we will see in Appendix \ref{sec.app3}, the
above definition does not lead to a finite operator.

\subsection{The renormalization scheme $[ ~\cdot~]_r^g$}

To obtain an operator that is demonstrably finite and that, for
simplicity's sake, can be obtained from the quadratic counter term alone, 
we will generalize equation \eq{eq.rtc.finiteness-from-OPE} to include
finite terms in the counterterm:
\be \label{eq.rtc.regulate-a-product}
\bignorder[g]{\prod_{i}V(t_{i})} \eqdef
\exp\left(-\frac{1}{2}\int dt_{1}dt_{2}~g(t_1, t_2)
\frac{\delta}{\delta V(t_{1})}\frac{\delta}{\delta V(t_{2})}\right)
\prod_{i}V(t_{i})~,
\ee 
where $g(t_1,t_2) = \frac{1}{(t_1-t_2)^2} + \mathrm{finite~terms}$.

We can rewrite  equation (\ref{eq.rtc.double-ren-defn}) using this new notation
 \be
 \left[(V(a,b))^{2}\right]_{r}=\int_{a}^{b}~dt_{1} dt_{2}
 ~\bignorder[g^D_{ab}]{ V(t_{1})V(t_{2})}~.
 \ee
To compare with equation (\ref{eq.rtc.double-ren-defn}) we notice that 
since the integrand in the above equation is finite, we can equivalently write
\begin{subequations}\begin{align}
 \left[(V(a,b))^{2}\right]_{r}&=
\lim_{\epsilon\rightarrow0}\int_{\Gamma_{\epsilon}^{a,b}(t_{1},t_{2})}dt_{1} dt_{2}
 ~\bignorder[g^D_{ab}]{ V(t_{1})V(t_{2})} \\ &=
\lim_{\epsilon\rightarrow0}\int_{\Gamma_{\epsilon}^{a,b}(t_{1},t_{2})}dt_{1} dt_{2}
 ~ \left [ V(t_{1})V(t_{2}) - g^D_{ab}(t_1,t_2) \right ]~,
 \end{align}\end{subequations}
which allows us to split the integral into two pieces (neither of which
is finite for $\epsilon \rightarrow 0$).

 The above equation introduces a new notation:
 $\Gamma^{a,b}_{\epsilon}(t_1, \ldots, t_n)
 := \{(t_1, \ldots, t_n) ~|~ a\leq t_i\leq b, |t_i-t_j|>\epsilon\}$.
 This is the same region of integration that is used
 for $(V(a,b)^{n})_{\epsilon}$.  For the sake of brevity, we often
 omit the list of parameters $(t_1, \ldots, t_n)$.  

Requiring that the function $g_{ab}^D$  not depend on $\epsilon$,  to match (\ref{eq.rtc.double-ren-defn})  
we must have
\begin{subequations}
 \be
 \int_{a}^{b-\epsilon}dt_{1}\int_{t_{1}+\epsilon}^{b}dt_{2}~g^{D}_{ab}(t_{1},t_{2})
 = G^D_{ab}+O(\epsilon)~,
\label{g-condition}
 \ee
 which can be satisfied by, for example 
 \be
\label{eq.rtc.all-ren-practical-D}
 g^{D}_{ab}(t_{1},t_{2})=\frac{1}{\left(t_{1}-t_{2}\right)^{2}}+\frac{2}{(b-a)^{2}}\left(1+\ln(b-a)+C_{0}+(b-a)C_{1}\right)~.
 \ee
\end{subequations}
We will be able to show shortly that details of the function
 $g_{ab}^D$ are not important as long as \eq{g-condition} is satisfied.
However, notice that the counterterm $\bignorder[g^D_{ab}]{~}$ does depend on $a$ and $b$: it is not
a `local' regulator like that in \eq{eq.rtc.finiteness-from-OPE}.

At higher orders, we now make the following definition
for a specific higher order regularization scheme:
 \be \label{eq.rtc.manyint-ren-little-g}
 \left[(V(a,b))^{n}\right]_{r}^g ~\eqdef~ 
\int_{a}^{b}~
dt_{1}\ldots dt_{n} \bignorder[g_{ab}^D]{V(t_1) \dots V(t_n)}~.
 \ee
The guiding principle of this scheme, which makes it easier
to prove that is satisfies all the required conditions, is to use 
the same integration region for every term related to a single 
 renormalized integrated operator. This requirement fixes
finite parts of the counterterms at higher order
in terms of those at quadratic orders so the only free
parameters are $C_0$ and $C_1$ (which enter through the specific 
counterterm $g$ we are using).  For example,
in the language of the previous section, we have 
$C_0^{(3)}=C_0^{(3),DL}=C_0^{(3),DR}=-(3+\ln 2) + C_0$.

The renormalization scheme has a simple form when applied to
an exponential:
\be \label{eq.rtc.highorder-one-factor-little-g}
\left[e^{\lambda V(a,b)}\right]^g_{r}  =
 \sum_{n=0}^\infty ~~ \int_{a}^{b}
\prod_{i=1}^n dt_i~~  e^{-\half \lambda^{2} g^{D}_{ab} } ~e^{\lambda V}~.
\ee
The notation here is similar to that commonly used for 
the Chern-Simmons action on a D-brane: under an $n$-dimensional
integral, we include all the terms from the Taylor expansion of the
integrand that have the right number of variables to saturate the integral.
It is easy to see that this is the same definition as that in equation
(\ref{eq.rtc.manyint-ren-little-g}).  

Expanding the above in powers of $\lambda$ we obtain a different
form
\be \label{aaaa} 
 \left[(V(a,b))^{n}\right]^g_{r}=\int_a^b dt_{1}\ldots dt_{n}
 \sum_{\sigma \in S_n} ~\sum_{0\leq k\leq\frac{n}{2}}\frac{(-1)^{k}}{2^{k}k!(n-2k)!}\prod_{i=1}^{k}
g^{D}_{ab}(t_{\sigma(2i-1)},t_{\sigma(2i)})\prod_{j=2k+1}^{n}V(t_{\sigma(j)})~.
\ee
Reinstating $\epsilon$ regularization allows us
to remove the cumbersome symmetrization sum from the above expression
 \be \label{bbbb}
 \left[(V(a,b))^{n}\right]^g_{r}=\lim_{\epsilon\rightarrow0}\int_{\Gamma_{\epsilon}^{a,b}(t_{1},\ldots,t_{n})}dt_{1}\ldots dt_{n}
 \sum_{0\leq k\leq\frac{n}{2}}\frac{(-1)^{k}n!}{2^{k}k!(n-2k)!}\prod_{i=1}^{k}g^{D}_{ab}(t_{2i-1},t_{2i})\prod_{j=2k+1}^{n}V(t_{j})~.
 \ee
In contrast to equation \eq{aaaa}, the integrand in equation \eq{bbbb} is not finite
and the integration region must be modified appropriately.

Now, consider a renormalization scheme with a different function 
$\tilde g^D_{ab} =  g^D_{ab}(t_1,t_2) + \Delta_{ab}^{D}(t_1,t_2)$ where the difference 
$\Delta_{ab}^{D}$ is assumed to be a finite  function of $t_1$ and $t_2$.  
As is shown in Appendix \ref{AA}, 
  \begin{subequations}
\label{g-independence}
\begin{align}
&[V(a,b)^{n}]_r^{\tilde g = g + \Delta} = \\ 
&\int_a^b dt_{1}\ldots dt_{n}
 \sum_{\sigma \in S_n} ~\sum_{0\leq k\leq\frac{n}{2}}\frac{(-1)^{k}}{2^{k}k!(n-2k)!}\prod_{i=1}^{k}
\tilde g^{D}_{ab}(t_{\sigma(2i-1)},t_{\sigma(2i)})\prod_{j=2k+1}^{n}V(t_{\sigma(j)}) \\
&=~~~\sum_{0\leq m \leq\frac{n}{2}}~ \frac{1}{m!} \left (-\half 
\int_a^b ds_{1} ds_{2} \Delta^D_{ab}(s_1,s_2) \right )^m ~~ [V(a,b)^{n-2m}]_r~.
 \end{align}\end{subequations}
This implies, in particular, that if $\int_a^b ds_{1} ds_{2} \Delta^D_{ab}(s_1,s_2) = 0$, 
then the operator renormalized using $\tilde g^D_{ab}$ is the
same as that renormalized using $g^D_{ab}$. However, 
if $\Delta^{ab} :=\int_a^b ds_{1} ds_{2} \Delta^D_{ab}(s_1,s_2)  \neq 0$, the new operator
is different, but the difference exponentiates:
\be
\label{cccc}
\left[e^{\lambda V(a,b)}\right]_{r}^{\tilde g = g + \Delta} = 
e^{-\half \lambda^2 \Delta_{ab}^{D}} \left[e^{\lambda V(a,b)}\right]_{r}^{g}~. 
\ee

\subsection{Renormalization of unintegrated operators}

Having defined a regularization scheme for $V(a,b)^n$, we now move
on to $V(a)V(a,b)^n$.  Again, we want to exponentiate our 
second-order scheme.  With a bit of abuse of notation, we
will use
\be
\bignorder[g_{ab}]{V(t_1) \ldots V(t_n)}
\ee
to mean an operator in which a pairwise divergence between any two
insertions is regulated by subtracting, as required, 
either $g^D_{ab}$,  $g^L_{ab}$ or  $g^R_{ab}$, where
\begin{subequations}
\be
\label{eq.rtc.all-ren-practical-L}
 g^{L}_{ab}(t)
\eqdef \frac{1}{\left(t-a\right)^{2}}+\frac{1}{\left(b-a\right)^{2}}+\frac{C_{L}}{b-a}~,
\ee
so that it satisfies
\be
 \int_{a+\epsilon}^{b}dt~g^{L}_{ab}(t)
=G^L+O(\epsilon)~,\label{eq.rtc.integrated-little-gL}
\ee
\end{subequations}
and where the definition of $g^R_{ab}$ follows along similar lines.
As was the case with $g^D_{ab}$, the exact form of the finite part of 
functions $g^L_{ab}$ and its counterpart $g^R_{ab}$ is not important, and
only its average value affects the operator.  We have used a
convenient and simple constant form in our definition above. 

So, for example, in the context of
\be
[V(a)V(a,b)^2]_r^{g} \eqdef \int_a^b~{dt_1 dt_2} \bignorder[g]{V(a) V(t_1) V(t_2)}~,
\ee
our notation $\bignorder[g]{~}$ indicates
\be
\bignorder[g]{V(a) V(t_1) V(t_2)} =  V(a)V(t_1)V(t_2) - V(a)g_{ab}^D(t_1,t_2) - 
 g_{ab}^L(t_1) V(t_2) -  g_{ab}^L(t_2) V(t_1) ~.
\ee

\subsection{Multiple regions of integration and replacement condition  \eq{eq.rtc.assume-replace} }

Using the exponential notation, we extend our definition
\eq{eq.rtc.highorder-one-factor-little-g} to more complicated
operators with several regions of integration  
\begin{eqnarray}  \label{eq.rtc.highorder-little-g-several-exponentials}
\left[\prod_{i=1}^{p}e^{\lambda_{i}V(a_{i},a_{i+1})}\right]^g_{r}
&\eqdef&~
\left ( \sum_{k_1=0}^\infty \ldots  \sum_{k_p=0}^\infty \right ) ~\prod_{i=1}^{p}\int_{a_i}^{a_{i+1}} d^{k_i}t 
\\ \nn &~&~~ \left(
\prod_{i=1}^{p}e^{-\half\lambda_{i}^{2}g^{D}_{a_{i},a_{i+1}}}~
\prod_{i=1}^{p-1}e^{-\half\lambda_{i}\lambda_{i+1}g^{E}_{a_{i},a_{i+1},a_{i+2}}}~
\prod_{i=1}^{p} e^{\lambda_{i}V_{a_i,a_{i+1}}}\right)_{\epsilon}~,
\end{eqnarray}
where we must define another counterterm function $g^E$:
\begin{subequations}
\be
\label{eq.rtc.all-ren-practical-E}
 g^{E}_{abc}(t_{1},t_{2},t_{3}) \eqdef 
\frac{1}{\left(t_{1}-t_{2}\right)^{2}}-\frac{1}{(c-b)(b-a)}\left(1+\ln\left(\frac{(c-b)(b-a)}{c-a}\right)+C_{0}\right)~,
\ee
to satisfy
\be
 \int_{a}^{b}dt_{1}\int_{\mathrm{max}(b,t_{1}+\epsilon)}^{c}dt_{2}~g^{E}_{abc}(t_{1},t_{2})~=~G^E_{abc}+O(\epsilon)~.\label{eq.rtc.integrated-little-gE}
\ee
\end{subequations}

In equation \eq{eq.rtc.highorder-little-g-several-exponentials}, 
all functions should be considered zero outside of their natural domain,
such as $(a_{i},a_{i+1})^2$ for $g^{D}_{a_{i},a_{i+1}}$.
To remove ambiguity, we have decorated $\lambda_i V$ with its
appropriate domain as well: $\lambda_i V_{a_i,a_{i+1}}$.  Finally, special attention
needs to paid to the domain of $g^{E}_{a_{i},a_{i+1},a_{i+2}}$.  
We have taken it to be the  region 
$(a,b)\times(b,c) \cup (b,c)\times(a,b)$, instead of 
$(a,b)\times(b,c)$.  This choice,
to double the domain of the function, which will be convenient below,
has resulted in a factor of 
$\half$ in the exponent containing $g^{E}_{a_{i},a_{i+1},a_{i+2}}$.

To verify that our renormalization scheme satisfies the replacement condition, 
we write a simpler version
of (\ref{eq.rtc.highorder-little-g-several-exponentials}) with only
two exponentials and equal couplings $\lambda_1=\lambda_2=\lambda$:
\begin{align}  \label{eq.rtc.highorder-little-g-two-exponentials}
\left[e^{\lambda V(a,b)} e^{\lambda V(b,c)}\right]_{r}
&=\lim_{\epsilon\rightarrow0}~  \sum_{k=0}^\infty \sum_{j=0}^\infty~ \int_{a}^{b} d^{k }t  \int_{b}^{c} d^{j} t
\left( e^{-\half\lambda^{2}g^{D}_{ab} -\half\lambda^{2}g^{D}_{bc} -\half\lambda^2 g^{E}_{abc}}~
e^{\lambda V_{a,b}}e^{\lambda V_{b,c}} \right)_{\epsilon}~.
\end{align}
We can now prove replacement in exponential notation: 
the $\epsilon$-map $(\ldots) \rightarrow (\ldots)_\epsilon$ is linear,
so the above equation implies that
\be
\left[e^{\lambda V(a,b)} e^{\lambda V(b,c)}\right]_{r}
=\left[e^{\lambda V(a,c)}\right]_{r}^{\tilde g^D_{ac} =  g^D_{ab} + g^D_{bc} + g^E_{abc}}~.
\ee
From our previous discussion, we have
\be
\left[e^{\lambda V(a,c)}\right]_{r}^{\tilde g^D_{ac} =  g^D_{ab} + g^D_{bc} + g^E_{abc}}~=
\left[e^{\lambda V(a,c)}\right]_{r}^{g^D_{ac}}~,
\ee
as long as $g^D_{ac} -(  g^D_{ab} + g^D_{bc} + g^E_{abc})$ is a finite function
and $\int_a^c d^t  g^D_{ac} -(  g^D_{ab} + g^D_{bc} + g^E_{abc})= 0$.
That $g^D_{ac} -(  g^D_{ab} + g^D_{bc} + g^E_{abc})$ is finite is obvious from
equations (\ref{eq.rtc.all-ren-practical-D}) and (\ref{eq.rtc.all-ren-practical-E})
when keeping in mind that the
union of natural domains of $g^D_{ab}$, $g^D_{bc}$ and $g^E_{abc}$
is the same as the domain of $g^D_{ac}$.  Further,
$\int_a^c d^2 t \left ( g^D_{ac} -(  g^D_{ab} + g^D_{bc} + g^E_{abc})
\right )= G^D_{ac} - (  G^D_{ab} + G^D_{bc} + G^E_{abc})$,
which vanishes if $C_0 = C^E$, the same condition we obtained
from the replacement condition at second order.

\subsection{Assumptions   \eq{eq.rtc.assume-factor}, \eq{eq.rtc.assume-reflect} and \eq{eq.rtc.assume-local}
} \label{sec.rtc.assume-ef}

To start with, we notice that the factorization assumption
\eq{eq.rtc.assume-factor} 
follows quite obviously from our renormalization scheme:
renormalized operators which are inserted away from each other
do not undergo any further renormalization when combined.

The assumption \eq{eq.rtc.assume-reflect} is also fairly 
straightforward to 
verify.  Examining equation 
\eq{eq.rtc.highorder-one-factor-little-g}
we see that assumption \eq{eq.rtc.assume-reflect} is satisfied because the 
region of integration relevant to $\left(V(a,b)^{n}\right)_{\epsilon}$,
parametrized by $t_1, \ldots, t_n$, is invariant under the map
$t_i \rightarrow (a+b)-t_i$ and because 
$g^{D}_{ab}(t_1,t_2) = g^{D}_{ab}(a+b-t_1,a+b-t_2)$.

The last assumption, \eq{eq.rtc.assume-local}, is trivial in our 
construction, since at no 
point in the renormalization of the integrated operators have we 
considered the wedge state on which they are embedded.  By 
constructing the counterterms using the local OPE rather than the 
two-point functions, we have avoided any difficulties that this 
assumption may have caused.  It is here that our approach differs
from that in work \cite{Kiermaier:2007vu}. 

\subsection{The first BRST condition \eq{eq.rtc.assume-Q1}}
\label{sec.rtc.assume-Q1}

Our proof follows that in \cite{Kiermaier:2007vu} quite closely,
while filling in some missing technical steps.
We  present it here in detail for completeness and to highlight where
our lemma \eq{eq.rtc.Q1-factorization-lemma} comes in.

The renormalized operator we start with this time is, as in 
\eq{eq.rtc.manyint-ren-little-g}, 
\be
\frac{1}{n!}\left[(V(a,b))^{n}\right]_{r}^{g}=
\sum_{k=0}^{\fln}\frac{(-1)^{k}}{2^{k}k!(n-2k)!}
\int_{\Gamma_{\epsilon}^{a,b}(t_{1},\ldots,t_{n})}dt_{1}\ldots dt_{n}
\prod_{i=1}^{k}g(t_{i},t_{i+k})\prod_{j=2k+1}^{n}V(t_{j})~.
\ee
The limit $\epsilon\rightarrow0$ will be implied throughout, but not
stated explicitly.  Also, because the counterterm 
$g^{D}_{ab}$ appears very frequently, we will drop the indices 
and simply refer to it as $g$ when this does not 
result in ambiguity.

We wish to show that 
\be
\frac{1}{n!}Q_{B}\left[V(a,b)^{n}\right]_{r}^{g}=\sum_{l=1}^{2}\frac{1}{(n-l)!}\left(\left[V(a,b)^{n-l}O_{R}^{(l)}(b)\right]_{r}^{g}-\left[O_{L}^{(l)}(a)V(a,b)^{n-l}\right]_{r}^{g}\right)~.
\ee

The BRST $Q_B$ operator acts like a derivative on the marginal operators $V$
(see equation \eq{eq.rtc.Q-primitives}), but not the counterterms $g$.  If the BRST operator
acted on \emph{both}, then its action on the 
renormalized operator would naturally contain
complete total derivatives and the proof of the BRST
condition would be simple.  Since it does not, we
effectively proceed as if it did and then subtract the
unnecessary extra terms this generates.  To do so,
we need to give a precise implementation of the 
morally correct statement that
\be \label{eq.rtc.schematic-Q1-splitting}
\left[V(a,b)^{n}\right]_{r} \leftrightarrow \left[V(a,b)^{n-1}\right]_{r}V(a,b)-(n-1)\left[V(a,b)^{n-2}\right]_{r}G^{D}_{ab}~.
\ee
We will achieve this with an add-and-subtract trick.  
The final result of this lengthy calculation is presented in 
equations (\ref{BRST1}) and (\ref{BRST1-Os}).

To begin, we use the action of the BRST operator on 
the marginal deformation \eq{eq.rtc.Q-primitives}
\begin{multline}
\frac{1}{n!}Q_{B}\left[V(a,b)^{n}\right]^g_{r}~=\\
\sum_{k=0}^{\fln}\frac{(-1)^{k}}{2^{k}k!(n-2k)!}\int_{\Gamma_{\epsilon}^{a,b}}d^{n}t~\partial_{t_{n}}\left((n-2k)\prod_{i=1}^{k}g(t_{i},t_{i+k})\prod_{j=2k+1}^{n}V(t_{j})c(t_{n})\right)\label{eq.rtc.Q1-beforesplit}
\end{multline}
We have left the term $(n-2k)$ explicit (instead of canceling it against
the exponential) so that the sum could be extended to $k=\fln$ for $n$ even.

We  now add and subtract the following quantity:  
\begin{subequations}\begin{gather}
\shoveleft{\sum_{k=1}^{\fln}\frac{(-1)^{k}}{2^{k}k!(n-2k)!}\int_{\Gamma_{\epsilon}^{a,b}}d^{n}t~\partial_{t_{n}}\left(2k\prod_{j=1}^{n-2k}V(t_{j})\prod_{i=n-2k+1}^{n-k}g(t_{i},t_{i+k})c(t_{n})\right)}\label{eq.rtc.Q1-simpleidentityA}\\
\shoveright{=-\sum_{k=0}^{\fln-1}\frac{(-1)^{k}}{2^{k}k!(n-2k-2)!}\int_{\Gamma_{\epsilon}^{a,b}}d^{n}t~\partial_{t_{n}}\left(\prod_{i=1}^{k}g(t_{i},t_{i+k})\prod_{j=2k+1}^{n-2}V(t_{j})g(t_{n-1},t_{n})c(t_{n})\right)}\label{eq.rtc.Q1-simpleidentityC}
\end{gather}\end{subequations}
In going between the two lines, we have shifted the range of $k$,
cancelled a factor of $2k$ against the combinatorial factor in front
and relabeled the integration variables $t_i$ for $i<n$.
Now we take \eq{eq.rtc.Q1-beforesplit} and we add 
\eq{eq.rtc.Q1-simpleidentityA} and subtract 
\eq{eq.rtc.Q1-simpleidentityC}.  This gives us 
\be
\frac{1}{n!}Q_{B}\left[V(a,b)^{n}\right]_{r}~=~ \mathcal{A} ~+~\mathcal{B},
\ee
where
\begin{subequations}\label{eq.rtc.Q1-splitAB}\begin{align}
\begin{split}\mathcal{A}&\eqdef\int_{\Gamma_{\epsilon}^{a,b}}d^{n}t~
\partial_{t_{n}}\left(c(t_{n})\sum_{k=0}^{\fln}\frac{(-1)^{k}}{2^{k}k!(n-2k)!}\right .\\
&\qquad\qquad\left .\left [(n-2k)\prod_{i=1}^{k}g(t_{i},t_{i+k})\prod_{j=2k+1}^{n}V(t_{j})+2k\prod_{j=1}^{n-2k}V(t_{j})\prod_{i=n-2k+1}^{n-k}g(t_{i},t_{i+k})\right]\right)\end{split}\label{eq.rtc.Q1-splitA}\\
\mathcal{B}&\eqdef\sum_{k=0}^{\fln-1}\frac{(-1)^{k}}{2^{k}k!(n-2k-2)!}\int_{\Gamma_{\epsilon}^{a,b}}d^{n}t~\partial_{t_{n}}\left(\prod_{i=1}^{k}g(t_{i},t_{i+k})\prod_{j=2k+1}^{n-2}V(t_{j})g(t_{n-1},t_{n})c(t_{n})\right)\label{eq.rtc.Q1-splitB}
\end{align}\end{subequations}

To evaluate $\mathcal{A}$, we observe that if we symmetrize
its integrand over the variables $t_{1},\ldots,t_{n-1}$,
it will take on the  form
\be
\partial_{t_{n}}\left(c(t_{n})f(\vec{t})\right)~,
\ee
where
\be
f(\vec{t})=\frac{1}{(n-1)!}\sum_{u\epsilon S_{n}}\sum_{k=0}^{\fln}\frac{(-1)^{k}}{2^{k}k!(n-2k)!}\prod_{i=1}^{k}g(t_{u(i)},t_{u(i+k)})\prod_{j=2k+1}^{n}V(t_{u(j)})~.
\ee
In this form, $\mathcal{A}$ is completely symmetric in all $n$ variables $t_i$, except for
the factor of $c(t_n)$.  As we already showed when demonstrating
the finiteness of the renormalization scheme, this integrand
is completely finite.  In this symmetrized form, 
it is safe to change the integration region to $(a,b)^{n}$ and 
perform the (trivial) integral over $t_{n}$ using the fundamental 
theorem of calculus.  We can then change the remaining $(n-1)$-dimensional
region of integration back to an $\epsilon$-regulated one
$\Gamma_{\epsilon}^{a+\epsilon,b-\epsilon}(t_{1},\ldots,t_{n-1})$.
Finally, we relabel the integration variables and obtain a 
simpler expression 
\begin{multline}
\mathcal{A}=\int_{\Gamma_{\epsilon}^{a+\epsilon,b-\epsilon}(t_{1},\ldots,t_{n-1})}d^{n-1}t\sum_{k=0}^{\fln}\frac{(-1)^{k}}{2^{k}k!(n-2k)!}\\
\qquad\qquad\left((n-2k)\prod_{i=1}^{k}g(t_{i},t_{i+k})\prod_{j=2k+1}^{n-1}V(t_{j})(cV(b)-cV(a))\right.\hfill\\
\left.+2k\prod_{j=1}^{n-2k}V(t_{j})\prod_{i=2-2k+1}^{n-k-1}g(t_{i},t_{i+k})\left(g^{D}_{ab}(t_{n-k},b)c(b)-g^{D}_{ab}(t_{n-k},a)c(a)\right)\right)~.
\end{multline}
The expression above has a form reminiscent of
$\left[V(a,b)^{n-1}(cV(b)-cV(a))\right]_{r}$, as required;
however, one more adjustment is necessary: in
$\left[V(a,b)^{n-1}(cV(b)-cV(a))\right]_{r}$, $c(a)g^{L}_{ab}(t_{i},a)$
and $c(b)g^{R}_{ab}(t_{i},b)$
should appear in the correct places, but in the expression above,
it is $c(a)g^{D}_{ab}(t_{i},a)$  and $c(b)g^{D}_{ab}(t_{i},b)$ 
that appear instead (we have restored the decorations on $g$ here
to make this more apparent).  Fortunately, the difference between
$g^{D}_{ab}(a,t_{i})$ and $g^{L}_{ab}(a,t_{i})$ is finite, so
we can write
\begin{multline}
\mathcal{A}~=~\frac{1}{(n-1)!}\left[V(a,b)^{n-1}(cV(b)-cV(a))\right]_{r}^g~+\\
\frac{1}{(n-2)!}\left[V(a,b)^{n-2}\right]_{r}^g\left(c(b)\int_{a}^{b}dt(g^{R}_{ab}(t,b)-g^{D}_{ab}(t,b))-c(a)\int_{a}^{b}dt(g^{L}_{ab}(a,t)-g^{D}_{ab}(a,t))\right)~.
\end{multline}
From the definitions in \eq{eq.rtc.all-ren-practical-D} and 
\eq{eq.rtc.all-ren-practical-L} we then have 
\begin{subequations}\ba
g^{L}_{ab}(x,y)-g^{D}_{ab}(x,y)&=\frac{1}{(b-a)^{2}}+\frac{C^{L}}{b-a}+\frac{2}{(b-a)^{2}}\left(1+\ln(b-a)+C_{0}+(b-a)C_{1}\right)\\
&=\frac{1}{(b-a)^{2}}+\frac{C^{L}}{b-a}+f^{D}_{ab}~,
\end{align}\end{subequations}
where 
$f^{D}_{ab}\eqdef\frac{2}{(b-a)^{2}}(1+\ln(b-a)+C_{0}+(b-a)C_{1})$ is 
the constant part of $g^{D}_{ab}$.  Thus 
\begin{multline}
\mathcal{A}=\frac{1}{(n-1)!}\left[V(a,b)^{n-1}(cV(b)-cV(a))\right]_{r}^g\\
+\frac{1}{(n-2)!}\left[V(a,b)^{n-2}\right]_{r}^g\left(\frac{c(b)-c(a)}{b-a}+(c(b)-c(a))C^{L}-(c(b)-c(a))(b-a)f^{D}_{ab}\right)~.
\label{piece-a}
\end{multline}

To evaluate $\mathcal{B}$, we notice 
that the integrand diverges whenever $t_{n-1}$ and $t_{n}$ approach 
each other, but not when these two variables approach any of the 
others.  This alone is not enough to factorize the region of 
integration, but with \eq{eq.rtc.schematic-Q1-splitting} in mind we 
notice that the rest of the integrand (including the sum and 
combinatorial factors) is what we would see for 
$\left[V(a,b)^{n-2}\right]_{r}^g$, so there are no divergences due to 
$t_{i}$ approaching any other $t_j$ as long as $i<n-1$.  In appendix \ref{lemma},
we show that 
\be \label{eq.rtc.Q1-factorization-lemma}
\int_{\Gamma_{\epsilon}^{ab}}d^{n}t\int_{\Gamma_{\epsilon}^{ab}}d^{2}s~f(\vec{t})\partial_{s_{2}}\left(g^{D}_{ab}(s_{1},s_{2})c(s_{2})\right)=\int_{\Gamma_{\epsilon}^{ab}}d^{n}t~d^{2}s~f(\vec{t})\partial_{s_{2}}\left(g^{D}_{ab}(s_{1},s_{2})c(s_{2})\right)
\ee
for any function $f(\vec{t})$ which is finite on $(a,b)^n$.  
Thus,  the domain of integration can be changed to
$\Gamma^{a,b}_\epsilon(t_{1},\dots,t_{n-2})~\times~\Gamma^{a,b}_\epsilon(t_{n-1},t_{n})$
and we evaluate the integrals with respect to $t_{n-1}$ and $t_{n}$:  
\begin{subequations}\begin{align}
\mathcal{B}&=\sum_{k=0}^{\fln-1}\frac{(-1)^{k}}{2^{k}k!(n-2k-2)!}\int_{\Gamma_{\epsilon}^{a,b}}d^{n}t~\partial_{t_{n}}\left(\prod_{i=1}^{k}g(t_{i},t_{i+k})\prod_{j=2k+1}^{n-2}V(t_{j})g(t_{n-1},t_{n})c(t_{n})\right)\\
\begin{split}&\quad=\sum_{k=0}^{\fln-1}\frac{(-1)^{k}}{2^{k}k!(n-2k-2)!}\left(\int_{\Gamma_{\epsilon}^{a,b}(t_{1},\ldots,t_{n-2})}d^{n-2}t\prod_{i=0}^{k}g(t_{i},t_{i+k})\prod_{j=2k+1}^{n-2}V(t_{j})\right)\\
	&\qquad\times\left(\int_{a}^{b-\epsilon}dt_{1}\int_{t_{1}+\epsilon}^{b}dt_{2}+\int_{a+\epsilon}^{b}dt_{1}\int_{a}^{t_{1}-\epsilon}dt_{2}\right)\partial_{t_{2}}\left(g(t_{1},t_{2})c(t_{2})\right)\end{split}\\
\begin{split}&\quad=\frac{\left[V(a,b)^{n-2}\right]_{r}^g}{(n-2)!}\left(\frac{c(b)-c(a)}{\epsilon}+\frac{c(a)-c(b)}{b-a}+\int_{a}^{b-\epsilon}dt\frac{c(t)-c(t+\epsilon)}{\epsilon^{2}}\right.\\
	&\qquad+\left.\vphantom{\frac{c(t)}{\epsilon}}(c(b)-c(a))(b-a)f^{D}_{ab}\right)\end{split}\\
&\quad=\frac{\left[V(a,b)^{n-2}\right]_{r}^g}{(n-2)!}\left(\frac{c(a)-c(b)}{b-a}+\frac{\partial c(b)}{2}+\frac{\partial c(a)}{2}+(c(b)-c(a))(b-a)f^{D}_{ab}\right)~.
\label{piece-b}
\end{align}\end{subequations}

Putting (\ref{piece-a}) and (\ref{piece-b}) together, 
several terms cancel and we get
\be
\begin{split}\mathcal{A}+\mathcal{B}
&=\frac{\left[V(a,b)^{n-1}cV(b)\right]_{r}^g}{(n-1)!}+\frac{\left[V(a,b)^{n-2}\right]_{r}^g}{(n-2)!}\left(\frac{\partial c(b)}{2}+C^{R}c(b)\right)\\
	&\qquad-\frac{\left[cV(a)V(a,b)^{n-1}\right]_{r}^g}{(n-1)!}+\left(\frac{\partial c(a)}{2}-C^{L}c(a)\right)\frac{\left[V(a,b)^{n-2}\right]_{r}^g}{(n-2)!}~.\end{split}\label{eq.rtc.Q-Vabn}
\ee
Multiplying this it by $\lambda^{n}$ and 
then summing over $n$, we arrive at the precise form we wanted:  
\be
Q_{B}\left[e^{\lambda V(a,b)}\right]_{r}^g
=\left[e^{\lambda V(a,b)}O_{R}(b)\right]_{r}^g
-\left[O_{L}(a)e^{\lambda V(a,b)}\right]_{r}^g~,
\label{BRST1}
\ee
where 
\be
O_{L}(a)=\lambda cV(a)-\frac{\lambda^{2}}{2}\partial c(a)+\lambda^{2}C^L c(a),\quad
O_{R}(b)=\lambda cV(b)+\frac{\lambda^{2}}{2}\partial c(b)+\lambda^{2}C^R c(b)~.
\label{BRST1-Os}
\ee
As has already been discussed, the explicit dependence of $O^L$ and $O^R$ on $C^L$ and 
$C^R$ is there to cancel the dependence of the renormalization scheme on these
parameters.  It can be shown that
\be
\left[V(a)e^{\lambda V(a,b)}\right]_r^{\tilde{g}^L=g^L+\Delta^L}
= [(V(a)e^{\lambda V(a,b)}]_r^g -
\left ( \lambda\int_a^b dt \Delta^{L}(t) \right ) \left[e^{\lambda V(a,b)}\right]_r^g,
\label{BRST1-shift}
\ee
which implies that the renormalized operators $\left[e^{\lambda V(a,b)}O_{R}(b)\right]_{r}^g$
and $\left[O_{L}(a)e^{\lambda V(a,b)}\right]_{r}^g$ are independent of the 
choice of $C^L$ and $C^R$.

\subsection{The second BRST condition \eq{eq.rtc.assume-Q2}}

To avoid clutter, in this section we will set $C^L$ and $C^R$ to zero.
As we have stressed so far, these constants are a matter of choice
and do not affect the eventual SFT solution.  

\subsubsection{A note on notation}

To prove the second BRST assumption, it will be useful to introduce
more flexible notation than what we have introduced so far.  In particular,
we have written in equation \eq{eq.rtc.manyint-ren-little-g}
\begin{subequations}
 \be
 \left[(V(a,b))^{n}\right]_{r}^g ~=~
\int_{a}^{b}~
dt_{1}\ldots dt_{n} \bignorder[g_{ab}^D]{V(t_1) \dots V(t_n)}~.
 \ee
To be more specific, we could have written
 \be
 \left[(V(a,b))^{n}\right]_{r}^{g^D_{ab}} ~=~
\int_{a}^{b}~
dt_{1}\ldots dt_{n} \bignorder[g_{ab}^D]{V(t_1) \dots V(t_n)}~.
 \ee
Such notation will allow us to use a counterterm $g^D_{ab}$ whose
parameters do not match the region of integration of $V$ exactly,
for example
 \be
\label{not1}
 \left[(V(a+\epsilon,b))^{n}\right]_{r}^{g^D_{ab}} ~=~
\int_{a+\epsilon}^{b}~
dt_{1}\ldots dt_{n} \bignorder[g_{ab}^D]{V(t_1) \dots V(t_n)}~.
 \ee
Further, since the counterterms $g^D$, $g^L$ and $g^R$ will need
to be modified  independently, we will use a notation
\be
 \left[ ~\cdot~ \right]_{r}^{g^D_{ab},g^{L/R}_{ab}}
\ee
\end{subequations}
to list the appropriate counterterms and (when necessary) their
parameters.
\\ \\

To verify the second BRST condition, we must compute
\be
\frac{Q_{B}}{(n-1)!}\left[cV(a)V(a,b)^{n-1}\right]_{r}+\frac{Q_{B}}{(n-2)!}\left(-\frac{1}{2}\partial c(a) \left[V(a,b)^{n-2}\right]_{r}\right)~.
\label{brst2-starting-point}
\ee
From the first BRST condition, the second term is 
\begin{multline}\label{eq.rtc.Q2-n-T45}
-\frac{Q_{B}}{(n-2)!}\frac{1}{2}\partial c(a)\left[V(a,b)^{n-2}\right]^{g}_r
=-\frac{1}{(n-2)!}\frac{c\partial^{2}c(a)}{2}\left[V(a,b)^{n-2}\right]_r^{g}\\
+\frac{1}{(n-3)!}\frac{\partial c(a)}{2}\left[V(a,b)^{n-3}cV(b)\right]_r^{g}+\frac{1}{(n-3)!}\frac{c\partial c(a)}{2}\left[V(a)V(a,b)^{n-3)}\right]_r^{g}\\
+\frac{1}{(n-4)!}\frac{\partial c(a)}{2}\left[V(a,b)^{n-4}\right]_r^{g}\frac{\partial c(b)}{2}~.
\end{multline}
In what follows, we need to know what happens when the BRST
operator acts on operators renormalized using a different
counterterm $\tilde g^D = g_{ab}^D + \Delta^D_{ab}$ instead of $g_{ab}^D$.
Recalling equation \eq{cccc}, we can write
\begin{subequations}\begin{align}
Q_{B}\left[e^{\lambda V(a,b)}\right]_r^{\tilde{g}_{ab}^D}
&=e^{-\frac{\lambda^{2}}{2}\int_{a}^{b}d^{2}s\,\Delta^{D}_{ab}(s_{1},s_{2})}Q_{B}\left[e^{\lambda V(a,b)}\right]_r^{g_{ab}^D}\\
&=e^{-\frac{\lambda^{2}}{2}\int_{a}^{b}d^{2}s\,\Delta^{D}_{ab}(s_{1},s_{2})}
\left(\left[e^{\lambda V(a,b)}O_{R}(b)\right]_r^{g_{ab}^D,g_{ab}^R}-\left[O_{L}(a)e^{\lambda V(a,b)}\right]_r^{g_{ab}^D,g_{ab}^L}\right)\\
&=\left[e^{\lambda V(a,b)}O_{R}(b)\right]_r^{\tilde{g}_{ab}^{D},g_{ab}^{R}}-\left[O_{L}(a)e^{\lambda V(a,b)}\right]_r^{\tilde{g}_{ab}^{D},g_{ab}^{L}}~,\\
\intertext{where  $O_{L/R}$ have the form given in equation
\eq{BRST1-Os}. To shift from $g_{ab}^{L/R}$ to $\tilde g_{ab}^{L/R} = g_{ab}^{L/R} + \Delta^{R/L}_{ab}$ we must
use equation \eq{BRST1-shift}: }
\begin{split}Q_{B}\left[e^{\lambda V(a,b)}\right]_r^{\tilde{g}_{ab}^D}
&=\left[e^{\lambda V(a,b)}\left(O_{R}(b)+\lambda^{2}\int_{a}^{b}dt\,\Delta^{R}_{ab}(t,b)\right)\right]_r^{\tilde{g}_{ab}^D,\tilde{g}_{ab}^{R}}\\
  &\qquad\qquad\qquad\qquad\qquad-\left[\left(O_{L}(a)+\lambda^{2}\int_{a}^{b}dt\,\Delta^{L}_{ab}(t)\right)e^{\lambda V(a,b)}\right]
_r^{\tilde{g}_{ab}^D,\tilde{g}_{ab}^{L}}\end{split}\\
\begin{split}&=\left[e^{\lambda V(a,b)}O_{R}(b)\right]_r^{\tilde{g}_{ab}^D,\tilde{g}_{ab}^{R}}
-\left[O_{L}(a)e^{\lambda V(a,b)}\right]_r^{\tilde{g}_{ab}^D,\tilde{g}_{ab}^{L}}\\
  &\qquad\qquad\qquad\qquad\qquad+\lambda^{2}\int_{a}^{b}dt\left(\Delta^{R}_{ab}(t,b)-\Delta^{L}_{ab}(t)\right)
\left[e^{\lambda V(a,b)}\right]_r^{\tilde{g}_{ab}^D}~.\end{split}\label{eq.rtc.Q1-tildeg}
\end{align}\end{subequations}
If $\Delta^{R}_{ab}=\Delta^{L}_{ab}$, we have that
\be
Q_{B}\left[e^{\lambda V(a,b)}\right]_r^{\tilde{g}_{ab}^D}
=\left[e^{\lambda V(a,b)}O_{R}(b)\right]_r^{\tilde{g}_{ab}^D,\tilde{g}_{ab}^{R}}
-\left[O_{L}(a)e^{\lambda V(a,b)}\right]_r^{\tilde{g}_{ab}^D,\tilde{g}_{ab}^{L}}~,
\label{eq.rtc.Q1-tildeg2}
\ee
i.e., the first BRST condition has the same form and uses the same
operators $O_{L/R}$ given in equation \eq{BRST1-Os} for any 
counterterms $g^D$ and $g^{L/R}$.

We will make use of this fact when $\tilde g$ is different from $g$
because it uses different values of $a$ and $b$ by a small amount $\epsilon$.  
For example, we might have $\tilde g^R_{ab} = g^R_{a+\epsilon,b}$ and
$\tilde g^L_{ab} = g^L_{a,b-\epsilon}$.  Then, since we are taking $C^{L/R}=0$ in this section,
$\Delta^L_{ab} = \Delta^R_{ab} = \frac{1}{(b-a+\epsilon)^{2}} -\frac{1}{(b-a)^{2}}$,
and we can use results \eq{BRST1} and \eq{BRST1-Os} without any changes.

With these preliminaries out of the way, the main part of the proof of 
\eq{eq.rtc.assume-Q2} consists of 
calculating the first term in \eq{brst2-starting-point}.
\begin{subequations}\be
\frac{Q_{B}}{(n-1)!}\left[cV(a)V(a,b)^{n-1}\right]_r^{g}=\frac{Q_{B}}{(n-1)!}\int_{a}^{b}d^{n-1}t\bignorder[g]{cV(a)\prod_{i=1}^{n-1}V(t_{i})}
\ee
At this point, we introduce a small parameter $\epsilon$
which is implicitly taken to zero.
Since the  integrand is  finite, we can modify the integration
region.  We make an $\epsilon$-sized modification to the integration
region at $a$ to examine the divergence there and write,
using notation \eq{not1}:
\be
\frac{Q_{B}}{(n-1)!}\int_{a+\epsilon}^{b}d^{n-1}t\left(cV(a)\bignorder[g_{ab}]{\prod_{i=1}^{n-1}V(t_{i})}-(n-1)c(a)g^{L}_{ab}(a,t_{1})\bignorder[g_{ab}]{\prod_{i=2}^{n-1}V(t_{i})}\right)~.
\ee
Using the fact that
$Q_B(cV)=0$ and then rewriting some operators in the
renormalized form with an understanding that the
implicit counterterm present in $[~]_r$ be taken to zero before 
$\epsilon$ gives
\begin{multline}\label{eq.rtc.Q2-n-T123-beforeQ}
-\frac{cV(a)}{(n-1)!}Q_{B}\left[V(a+\epsilon,b)^{n-1}\right]_r^{g_{ab}^D}+\frac{c(a)}{(n-2)!}\int_{a+\epsilon}^{b}dt\,g^{L}_{ab}(t)\,Q_{B}\left[V(a+\epsilon,b)^{n-2}\right]_r^{g_{ab}^D}\\
    -\frac{c\partial c(a)}{(n-2)!}\int_{a+\epsilon}^{b}dt\,g^{L}_{ab}(t)\,\left[V(a+\epsilon,b)^{n-2}\right]_r^{g_{ab}^D}~.
\end{multline}
The BRST operator can now act on these renormalized operators using 
\eq{eq.rtc.Q1-tildeg2} since the $\epsilon$-regulator is holding
the unintegrated insertion `away', resulting in:
\begin{multline}\label{eq.rtc.Q2-n-T123-afterQ}
    -\frac{cV(a)}{(n-2)!}\left(\left[V(a+\epsilon,b)^{n-2}cV(b)\right]_r^{g_{ab}^D,g_{ab}^R}
-\left[cV(a+\epsilon)V(a+\epsilon,b)^{n-2}\right]_r^{g_{ab}^D,g_{ab}^L}\right)\\
  -\frac{cV(a)}{(n-3)!}\left[V(a+\epsilon,b)^{n-3}\right]_r^{g_{ab}^D}\left(\frac{1}{2}\partial c(b)+\frac{1}{2}\partial c(a)\right)\\
  +\frac{c(a)}{(n-3)!}\int_{a+\epsilon}^{b}dt\,g^{L}_{ab}(t)\left(\left[V(a+\epsilon,b)^{n-3}cV(b)\right]_r^{g_{ab}^D,g_{ab}^R}
-\left[cV(a+\epsilon)V(a+\epsilon,b)^{n-3}\right]_r^{g_{ab}^D,g^L_{ab}}\right)\\
  +\frac{c(a)}{(n-4)!}\int_{a+\epsilon}^{b}dt\,g^{L}_{ab}(t)\,\left[V(a+\epsilon,b)^{n-4}\right]_r^{g^D_{ab}}\left(\frac{1}{2}\partial c(b)+\frac{1}{2}\partial c(a)\right)\\
  -\frac{c\partial c(a)}{(n-2)!}\int_{a+\epsilon}^{b}dt\,g^{L}_{ab}(t)\,\left[V(a+\epsilon,b)^{n-2}\right]_r^{g^D_{ab}}~.
\end{multline}
Rearranging and recombining some of the integrands into finite combinations, and
in one place using the fact that $\int_{a+\epsilon}^b dt~ g^L_{ab}(t) = \epsilon^{-1} + {\cal O}(\epsilon)$, we get
\begin{align}\label{eq.rtc.Q2-n-beforecombinestep}
\begin{split}
    &-\frac{1}{(n-2)!}\int_{a+\epsilon}^{b}d^{n-2}t\bignorder[g_{ab}]{cV(a)\prod_{i=1}^{n-2}V(t_{i})cV(b)}\\
  &\qquad-\frac{1}{(n-3)!}\int_{a+\epsilon}^{b}d^{n-3}t\bignorder[g_{ab}]{cV(a)\prod_{i=1}^{n-3}V(t_{i})}\frac{1}{2}\partial c(b)\\
  &\qquad-\frac{1}{(n-3)!}\int_{a+\epsilon}^{b}d^{n-3}t\frac{1}{2}c\partial c(a)\bignorder[g_{ab}]{V(a)\prod_{i=1}^{n-3}V(t_{i})}\\
  &\qquad+\frac{1}{(n-2)!}\int_{a+\epsilon}^{b}d^{n-2}t\left(cV(a)\norder[g_{ab}]{cV(a+\epsilon)\prod_{i=1}^{n-2}V(t_{i})}-\frac{c\partial c(a)}{\epsilon}\bignorder[g_{ab}]{\prod_{i=1}^{n-2}V(t_{i})}\right.\\
  &\qquad\qquad\qquad\qquad\qquad\qquad\left.-(n-2)c(a)g^{L}_{ab}(a,t_{1})\bignorder[g_{ab}]{cV(a+\epsilon)\prod_{i=2}^{n-2}V(t_{i})}\right)~.
\end{split}\end{align}
Where the integrands are finite, we can now remove the $\epsilon$-regulator on the
integration region.  The resulting expression is
\begin{multline}\label{eq.rtc.Q2-n-T123end}
    -\frac{1}{(n-2)!}\left[cV(a)V(a,b)^{n-2}cV(b)\right]_r^{g}-\frac{1}{(n-3)!}\left[cV(a)V(a,b)^{n-3}\right]_r^{g}\frac{1}{2}\partial c(b)\\
  -\frac{c\partial c(a)}{2(n-3)!}\left[V(a)V(a,b)^{n-3}\right]_r^{g}+\frac{c\partial^{2}c(a)}{2(n-2)!}\left[V(a,b)^{n-2}\right]_r^{g}~,
\end{multline}
\end{subequations}
where, to simplify the last and most complicated term in equation \eq{eq.rtc.Q2-n-beforecombinestep},
we have examined the following chain of equalities,\footnote
{Consistent with our notation, we include a counterterm for $V(a)V(a+\epsilon)$ in
$\norder[g_{ab}]{V(a)V(a+\epsilon)\ldots}$.  The finite part of this counterterm 
is irrelevant since the ghost factor will suppress it.} 
\begin{subequations}\begin{align}
\norder[g_{ab}]{cV(a)&cV(a+\epsilon)V(a+\epsilon,b)^{n-2}}\\
\begin{split}&=cV(a)\norder[g_{ab}]{cV(a+\epsilon)V(a+\epsilon,b)^{n-2}}-\frac{c(a)c(a+\epsilon)}{\epsilon^{2}}\norder[g_{ab}]{V(a+\epsilon,b)^{n-2}}\\
  &\qquad-(n-2)c(a)\int_{a+\epsilon}^{b}dt\,g^{L}_{ab}(t)\norder[g_{ab}]{cV(a+\epsilon,b)V(a+\epsilon,b)^{n-3}} + {\cal{O}}(\epsilon)\end{split}\\
\begin{split}&=cV(a)\norder[g_{ab}]{cV(a+\epsilon)V(a+\epsilon,b)^{n-2}}-\frac{c\partial c(a)}{\epsilon}\norder[g_{ab}]{V(a+\epsilon,b)^{n-2}}\\
  &\qquad-\frac{1}{2}c\partial^{2}c(a)\norder[g_{ab}]{V(a+\epsilon,b)^{n-2}}\\
  &\qquad-(n-2)c(a)\int_{a+\epsilon}^{b}dt\,g^{L}_{ab}(t)\norder[g_{ab}]{cV(a+\epsilon,b)V(a+\epsilon,b)^{n-3}}
 + {\cal{O}}(\epsilon)~.\end{split}
\end{align}\end{subequations}
These, together with the observation that
$\norder[g_{ab}]{cV(a)cV(a+\epsilon)V(a+\epsilon,b)^{n-2}}$ is of order $\epsilon$,
imply that we can replace the parentheses in \eq{eq.rtc.Q2-n-beforecombinestep} with 
$\frac{c\partial^{2}c(a)}{2}\norder[g_{ab}]{\prod_{i=1}^{n-2}V(t_{i})}$.

Parenthetically, it is worth noting that we can use explicit third order 
calculations to show that all of these steps are correct at that order. 
For example, at third order \eq{eq.rtc.Q2-n-T123-beforeQ} and 
\eq{eq.rtc.Q2-n-T123-afterQ} both match the explicitly computed 
third order result 
\be
\frac{1}{2}Q_{B}\left[cV(a)V(a,b)^{2}\right]_r^{g}=-\left[cV(a)V(a,b)cV(b)\right]_r^{g}-\frac{1}{2}cV(a)\partial c(b)+\frac{1}{2}c\partial^{2}c(a)V(a,b)-\frac{1}{2}c\partial cV(a)~.
\ee

Finally, we add the two pieces \eq{eq.rtc.Q2-n-T123end} and 
\eq{eq.rtc.Q2-n-T45} together to see  that
\begin{subequations}\begin{align}
\frac{Q_{B}}{(n-1)!}&\left[cV(a)V(a,b)^{n-1}\right]_r^{g}-\frac{Q_{B}}{(n-2)!}\frac{1}{2}\partial c(a)\left[V(a,b)^{n-2}\right]_r^{g}\\
\begin{split}&=-\frac{1}{(n-2)!}\left[cV(a)V(a,b)^{n-2}cV(b)\right]_r^{g}-\frac{1}{(n-3)!}\left[cV(a)V(a,b)^{n-3}\right]_r^{g}\frac{1}{2}\partial c(b)\\
  &\qquad-\frac{c\partial c(a)}{2(n-3)!}\left[V(a)V(a,b)^{n-3}\right]_r^{g}+\frac{c\partial^{2}c(a)}{2(n-2)!}\left[V(a,b)^{n-2}\right]_r^{g}\\
  &\qquad-\frac{c\partial^{2}c(a)}{2(n-2)!}\left[V(a,b)^{n-2}\right]_r^{g}+\frac{1}{(n-3)!}\frac{1}{2}\partial c(a)\left[V(a,b)^{n-3}cV(b)\right]_r^{g}\\
  &\qquad+\frac{c\partial c(a)}{2(n-3)!}\left[V(a)V(a,b)^{n-3}\right]_r^{g}+\frac{1}{(n-4)!}\frac{1}{2}\partial c(a)\left[V(a,b)^{n-4}\right]_r^{g}\frac{1}{2}\partial c(b)\end{split}\\
\begin{split}&=-\frac{1}{(n-2)!}\left[cV(a)V(a,b)^{n-2}cV(b)\right]_r^{g}-\frac{1}{(n-3)!}\left[cV(a)V(a,b)^{n-3}\right]_r^{g}\frac{1}{2}\partial c(b)\\
  &\qquad+\frac{1}{(n-3)!}\frac{1}{2}\partial c(a)\left[V(a,b)^{n-3}cV(b)\right]_r^{g}+\frac{1}{(n-4)!}\frac{1}{2}\partial c(a)\left[V(a,b)^{n-4}\right]_r^{g}\frac{1}{2}\partial c(b)~.\end{split}
\end{align}\end{subequations}
Summing this expression over $n$ gives 
\begin{multline}
Q_{B}\left[\left(\lambda cV(a)-\frac{\lambda^{2}}{2}\partial c(a)\right)e^{\lambda V(a,b)}\right]_r^{g}\\
=-\left[\left(\lambda cV(a)-\frac{\lambda^{2}}{2}\partial c(a)\right)e^{\lambda V(a,b)}\left(\lambda cV(b)+\frac{\lambda^{2}}{2}\partial c(b)\right)\right]_r^{g}~.
\end{multline}
This proves that the second BRST assumption \eq{eq.rtc.assume-Q2} holds 
in this particular renormalization scheme at all orders.

\section{Conclusions}
\label{sec.discussion}

In this section, we will discuss the effect that our
free parameters have on the corresponing SFT solution.
We will discuss first the effect of the already
explicitly identified free parameters, and then consider the
existence of other free parameters.

In sections \ref{sec.rtc.quadratic} and \ref{sec.rtc.cubic}, we have discussed
parameters such as $C_L$, $C_R$, $C_0^{(3),DL}$ and $C_0^{(3),DR}$
that affect the renormalization scheme in a relatively trivial way.
They change only the explicit form of $O_R$ and $O_L$ but do not
change the corresponding SFT solution.  In contrast,
parameters $C_1$, $C_0$ and $C_0^{(3)}$ appear at first glance
to affect the renormalization scheme and the SFT solution.  Let's examine
these in some detail.

From equation \eq{cccc}, we see that our free
parameters $C_0$ and $C_1$ are simply a rescaling of the 
renormalized operator:
\be
\label{rescaling}
\left [ e^{\lambda V(a,b)} \right ]_{r}^{g(C_0,C_1)} ~=~
e^{- \lambda^2 (C_1(b-a) + C_0)}~
\left [ e^{\lambda V(a,b)} \right ]_{r}^{g(C_0=0,C_1=0)} ~.
\ee
To understand whether this implies a change in the SFT
solution, we consider equations (3.11) and (3.12) of  \cite{Kiermaier:2007vu}:
\begin{subequations}
\be
U ~\equiv~\sum_{n=0}^n~\lambda^n{U^{(n)}}~,
\label{KO1}
\ee
where
\be
\langle \phi, U^{(n)} \rangle ~=~\frac{1}{n!}
\langle f \circ \phi, [V(1,n)^n]_r \rangle_{W_{n}}~.
\label{KO2}
\ee
\end{subequations}
This pair of equations defines a string field $U$ from which
the SFT solution of Kiermaier and Okawa is constructed.
Following the details of the construction, we see that
a rescaling of $U$ by a $\lambda$-dependent factor
changes $U$ and therefore has impact on the SFT solution. 
This is because, in equation \eq{KO2},
the interval on which $V$ is integrated is different at every order: $b-a = n-1$.  
With a $\lambda$-dependent rescaling factor, in the
resulting expression for $U$, the width of the integration interval will
no longer match the power to which $V$ is raised, and the 
final expression for $U$ will be different.
A numerical computation \cite{matt-numerics-unpublished}
indicates that indeed $C_0$ and $C_1$ do affect the
SFT solution.
We leave the question of whether SFT solutions given by different values 
of $C_1$ and $C_0$ are related by a gauge transformations to future work,
and offer only one more observation: introducing a
nonzero $C_1$ is the same as replacing $V(t)$ with  $V(t) - \lambda C_1$.

Notice that the rescaling \eq{rescaling} is
consistent with our comparison between equations \eq{conformal}
and \eq{derivative-answer}: if $C_1$ is not zero, the $\frac{\partial}{\partial a}$
derivative  in equation \eq{conformal} has an additional
term from the derivative acting on the rescaling factor 
$e^{- \lambda^2 (C_1(b-a))}$.  The apparent non-primarity
of the bcc operator seems to arise from this
rescaling of the renormalized operator.

Leaving now the confines of the renormalization scheme defined
in section \ref{sec.rtc.all-orders}, we can ask whether changing
$C_0^{(3)}$ changes the SFT solution.  It is easy to see that generalizing
the rescaling in equation \eq{rescaling} to  include higher order terms,
as in
\be
\left [ e^{\lambda V(a,b)} \right ]_{\tilde r} ~=~
e^{-(C_0\lambda^2 + C_0^{(4)}\lambda^4 + \ldots)~ - ~(C_1\lambda^2 + C_1^{(4)}\lambda^4 + \ldots)(b-a)}
\left [ e^{\lambda V(a,b)} \right ]_{r}^{g(C_0=0,C_1=0)} ~,
\label{rescaling2}
\ee
does not result in a change of $C_0^{(3)}$ from the value
it has in the scheme of section  \ref{sec.rtc.all-orders},
$C_0^{(3)} = -(3+\ln 2) + C_0$.  There is, however, 
another simple change in the renormalization schemes which does
affect $C_0^{(3)}$: a renormalization of the perturbation parameter $\lambda$.
Specifically, we can take
\be
\left [ e^{\lambda V(a,b)} \right ]_{\tilde r} ~=~
\left [ e^{(\lambda + 6 \Delta C^{(3)}_0 \lambda^3 + \ldots) V(a,b)} \right ]_{r}^g~,
\label{rescaling3}
\ee
where $\Delta C^{(3)}_0 = C^{(3)}_0 + (3+\ln 2) - C_0$.
The conclusion is then that changing $C_0^{(3)}$ away from 
$-(3+\ln 2) + C_0$ does affect the SFT solution, but in a 
benign and easy to understand way: by reparametrizing the deformation flow.
This observation also explains why there is no independent 
parameter $C_1^{(3)}$.

As equation \eq{rescaling2}  makes clear, at higher orders
there are more parameters that will affect the SFT solution beyond
a reparametrization of the deformation flow.  The
first two of these are $C_0^{(4)}$ and $C_1^{(4)}$.  Are there any other,
more complicated modifications to the renormalization scheme
that affect the SFT solution and are not just a reparameterization of the flow?
To answer this question, we could, for example, repeat
the analysis of section \ref{sec.rtc.cubic} at quartic order in $\lambda$
(to see whether there are any parameters other than $C_0^{(4)}$ and $C_1^{(4)}$).
This, however, is complicated. Not only are there more terms, but constructing
the most general finite renormalization scheme at this order is nontrivial:
recall that the naive guess in equation \eq{eq.rtc.highorder-one-factor}
turned out to not be finite (see Appendix \ref{sec.app3} for details).  
We are not able to offer an analysis beyond third order here, but do briefly
discuss a possible approach in the following subsection.

\subsection{Renormalization operator}

A good renormalization scheme must make the operator $e^{\lambda V(a,b)}$
finite and satisfy the conditions \eq{eq.rtc.assume-all}.  In our analysis
in sections \ref{sec.rtc.quadratic} and \ref{sec.rtc.cubic}, we saw
that the conditions of factorization \eq{eq.rtc.assume-factor}
and replacement \eq{eq.rtc.assume-replace} place strong constraints on 
possible renormalization parameters.  Since we have already identified the
replacement condition as essentially a linearity condition, to
get this condition `for free', we could implement our renormalization
scheme as a linear operator.  This approach will require the 
`extended linearity' of \eq{linearity-1}, and so it will produce 
restrictions such as $C^{L}=C^{R}=C_{1}$, which we will assume for 
this subsection. 

Consider then an operator $L_\epsilon$ given by
\be\begin{split}
L_\epsilon
&=\int dx dy ~\delta(x-y)G^{L}\frac{\delta}{\delta V(x)} \frac{\delta}{\delta V(y)} \\
&\quad+\frac{1}{2}\lim_{\Delta\rightarrow 0}\int dx dy ~\left(\delta^{\prime}(x-y+\Delta)-\delta^{\prime}(x-y-\Delta)\right)
G^{E}\frac{\delta}{\delta V(x)}\frac{\delta}{\delta 
V(y)}~.\end{split}
\ee
It has the property that
\begin{subequations}\begin{align}
L_\epsilon ~ V(a) V(a,b) &= G^L~, \\
L_\epsilon~ V(a,b)^2 &=  2(-G^{E}+(b-a)G^{L}) = 2 G^{D}_{ab}~, \\
L_\epsilon~V(a,b)V(b,c) &= G^{E}~, 
\end{align}\end{subequations}
thus it correctly produces the counterterms at quadratic order.

We could then ask whether, for any operator $A$ built out of 
integrated or fixed insertions of the marginal operator $V$, we should
define 
\be
[A]_r \overset{?}{=} \lim_{\epsilon \rightarrow 0}  (e^{-L_\epsilon} A)_\epsilon~.
\label{linear}
\ee
The answer is no: this would be equivalent to using equation \eq{eq.rtc.highorder-one-factor},
which we know not to be finite.  However, we might be able to `patch up' this problem
(and introduce more free parameters at the same time) by using a more general operator $L_\epsilon$.  Consider,
for example
\be
\label{eq.rtc.L-general-form}
L_\epsilon=\sum_{n=2}^{\infty}\int d^{n}x~\mathcal{L}^{(n)}_\epsilon(x_{1},\ldots,x_{n})\prod_{i=1}^{n}\frac{\delta}{\delta V(x_{i})}~.
\ee
This gives us a parametrization of sorts of possible renormalization
schemes at different orders. At the quadratic order, we have
\be
\mathcal{L}^{(2)}_\epsilon(x,y)
=G^{L}\delta(x-y)+\frac{1}{2}\lim_{\Delta\rightarrow0}G^{E}\left(\delta^{\prime}(x-y+\Delta)-\delta^{\prime}(x-y-\Delta)\right)~,
\ee
and at third order we could have
\be\label{eq.rtc.L3-ansatz}
\mathcal{L}^{(3)}_\epsilon(x,y,z)=A^{(3)}V(x)\delta(x-y)\delta(x-z)+B^{(3)}V(x)\delta(x-y)\delta(y-z)~,
\ee
where the free parameters uncovered in section \ref{sec.rtc.cubic}
are shown, by an explicit calculation, to be reproduced with
\be
C_{0}^{(3)}=C_{0}+A^{(3)}+B^{(3)},\quad 
C_{0}^{(3),DL}=C_{0}^{(3),DR}=C_{0}+A^{(3)}+\frac{7}{8}B^{(3)}~.
\ee

Since renormalization arises through an action of an operator here,
it is naturally linear, so the replacement condition \eq{eq.rtc.assume-replace}
would be naturally satisfied.
If we want to satisfy the factorization condition \eq{eq.rtc.assume-factor}, 
we just need some strategically placed $\delta$-functions,
as is explicit in equation \eq{eq.rtc.L3-ansatz}.

To extend this analysis to the next (quartic) order, we must account for 
subleading  divergences at fourth order that were uncovered in Appendix \ref{sec.app3}.
An explicit calculation gives additional divergent counterterms at fourth order:
\be
\mathcal{L}_{4}^\epsilon(x_{1},x_{2},x_{3},x_{4})\sim\frac{1}{\epsilon}\delta(x_{1}-x_{2})\delta(x_{2}-x_{3})\delta(x_{3}-x_{4})+O(\ln\epsilon)~.
\ee
Finite terms are of course allowed as well, and will contribute 
additional free parameters.  While we have not demonstrated that our scheme $[~~]_r^g$ is
of this type, we believe this to be true.

With this approach, we could in principle write down the most general finite scheme
at quartic order that satisfies conditions \eq{eq.rtc.assume-replace} and \eq{eq.rtc.assume-factor}.
Then, we would need to check that the BRST conditions do not impose any extra
restriction on the free parameters.  This would allow us to discover whether
there are any free parameters at quartic order that affect the SFT solution in a nontrivial way,
without analyzing all possible restrictions due to the replacement condition
at this order.

\appendix

\section{Comments on equation (\ref{eq.rtc.highorder-one-factor})}
\label{sec.app3}

While the renormalization schemes 
\be \tag{\ref{eq.rtc.highorder-one-factor}}
\lim_{\epsilon\rightarrow0}~ 
e^{-\lambda^{2} G^{D}_{a,b}}~\left(e^{\lambda V(a,b)}\right)_{\epsilon}
\ee
and
\be \tag{\ref{eq.rtc.highorder-one-factor-little-g}}
\left[e^{\lambda V(a,b)}\right]^g_{r}  =
 \sum_{n=0}^\infty ~~ \int_{a}^{b}
\prod_{i=1}^n dt_i~~  e^{-\half \lambda^{2} g^{D}_{ab} } ~e^{\lambda V}
\ee
look very similar, they are not equivalent.  The scheme we have been 
using, \eq{eq.rtc.highorder-one-factor-little-g}, is given at order 
$\lambda^{n}$ by \eq{bbbb}, and \eq{eq.rtc.highorder-one-factor} is 
similarly written out as 
 \be
  \label{eq.rtc.highorder-one-factor-expanded}
\lim_{\epsilon\rightarrow0}
 \sum_{0\leq k\leq\frac{n}{2}}\frac{(-1)^{k}n!}{k!(n-2k)!}
 (G^{D}_{ab})^k
 \int_{\Gamma_{\epsilon}^{a,b}(t_{2k+1},\ldots,t_{n})}dt_{1}\ldots dt_{n}
 \prod_{j=2k+1}^{n}V(t_{j}) ~.
 \ee

  The critical difference 
 between these two schemes is illustrated by 
 \begin{subequations}\ba 
(2G_{ab}^{D})^{k}&\int_{\Gamma_{\epsilon}^{a,b}(t_{1},\ldots,t_{i})}dt_{1}\ldots dt_{i}\prod_{j=1}^{i}V(t_{j})\\
&=\int_{\Gamma_{\epsilon}^{a,b}(t_{1},\ldots,t_{i})\times\Gamma_{\epsilon}^{a,b}(s_{1},s_{2})\times\ldots\times\Gamma_{\epsilon}^{a,b}(s_{2k-1},s_{2k})}dt_{1}\ldots dt_{i}~ds_{1}\ldots ds_{2k}\prod_{j=1}^{i}V(t_{j})\prod_{j=1}^{k}g(s_{2j-1},s_{2j})\\
&\neq\int_{\Gamma_{\epsilon}^{a,b}(t_{1},\ldots,t_{i},s_{1},\ldots,s_{2k})}dt_{1}\ldots dt_{i}ds_{1}\ldots ds_{2k}\prod_{j=1}^{i}V(t_{j})\prod_{j=1}^{k}g(s_{2j-1},s_{2j})~. \label{eq.rtc.ren-defn-KOform}
\end{align}\end{subequations}
We might try to argue that since the integrand has no singularity 
where one of the $s_{j}$ approaches a $t_{j}$ or an $s_{j}$ belonging 
to another counterterm, the difference vanishes as 
$\epsilon\rightarrow0$ and the difference between the integration 
regions shrinks.  The flaw in this reasoning is that when, for 
example, $s_{1}$ is close to one of the  $t_j's$, then the 
integrand does become large for $|s_{2}-t_j|<\epsilon$, 
an integration region which is included in one case but not
in the other.  As a  concrete example, at third order it can be shown 
that
\be \label{eq.rtc.factorizing-region-shift}
\left(\int_{a}^{b}dt\int_{\Gamma_{\epsilon}^{a,b}(s_{1},s_{2})}ds_{1}ds_{2}-\int_{\Gamma_{\epsilon}^{a,b}(t,s_{1},s_{2})}dtds_{1}ds_{2}\right)f(t)g(s_{1},s_{2})=\left(6+2\ln2\right)\int_{a}^{b}dt~f(t)~.
\ee
At fourth order, the problem becomes worse. Examining the difference
between the two renormalization schemes at this order, we see that
\begin{eqnarray}
\frac{1}{24}[V(a,b)^4]_r^g &=& 
 \frac{1}{24} \left ( V(a,b)^4 \right )_\epsilon
-\frac{1}{2}  \left ( V(a,b)^4 \right )_\epsilon G_{ab}^D
+\frac{1}{2}   \left (G_{ab}^D \right)^2 \\ \nn
&+&\frac{1}{4} \left ( \int_{\Gamma^{a,b}_\epsilon(t_1,t_2)} dt_1 dt_2 \int_{\Gamma^{a,b}_\epsilon(s_1,s_2)}  ds_1 ds_2 -
\int_{\Gamma^{a,b}_\epsilon(t_1,t_2,s_1,s_2)} dt_1 dt_2 ds_1 ds_2 
\right )\\& &~~~~~~~~~~~~~~~~~~~~~~~~~~~~~~~~~~~~~~ \left ( V(t_1)V(t_2) - \half g(t_1,t_2) \right )g(s_1,s_2) ~. \nn
\end{eqnarray}
The term on the last line is not finite as $t_1$ approaches $t_2$ and so the difference
between the two renormalization schemes is not finite as $\epsilon$ approaches zero.
Since $[V(a,b)^4]_r^g $ is demonstratively finite, it must be that
$ \frac{1}{24} \left ( V(a,b)^4 \right )_\epsilon
-\frac{1}{2}  \left ( V(a,b)^4 \right )_\epsilon G_{ab}^D
+\frac{1}{2}   \left (G_{ab}^D \right)^2$ is not.

\section{Proof of equation (\ref{g-independence})}
\label{AA}

We will explicitly write out the operator 
$\left[V(a,b)^{n}\right]_{r}^{\tilde{g}=g+\Delta}$ in order to 
compare it to the same operator renormalized with $g$:
  \begin{subequations}\begin{align}
&\int_a^b dt_{1}\ldots dt_{n}
 \sum_{\sigma \in S_n} ~\sum_{0\leq k\leq\frac{n}{2}}\frac{(-1)^{k}}{2^{k}k!(n-2k)!}\prod_{i=1}^{k}
\tilde g^{D}_{ab}(t_{\sigma(2i-1)},t_{\sigma(2i)})\prod_{j=2k+1}^{n}V(t_{\sigma(j)}) \\
\begin{split}& =\int_a^b dt_{1}\ldots dt_{n}
 \sum_{\sigma \in S_n} ~\sum_{0\leq k\leq\frac{n}{2}} 
~\frac{(-1)^{k}}{2^{k}k!(n-2k)!} \\
& ~~~~~ ~\sum_{m=0}^k~ {k \choose m}   \prod_{l=1}^{m} \Delta^D_{ab}(t_{\sigma(2l-1)},t_{\sigma(2l)})  \prod_{i=m+1}^{k} 
 g^{D}_{ab}(t_{\sigma(2i-1)},t_{\sigma(2i)})\prod_{j=2k+1}^{n}V(t_{\sigma(j)}) \end{split}\\
\begin{split}& =\int_a^b dt_{1}\ldots dt_{n}
 \sum_{\sigma \in S_n} ~\sum_{0\leq m \leq\frac{n}{2}}~\prod_{l=1}^{m} \frac{(-1)^m}{2^m m!}
\Delta^D_{ab}(t_{\sigma(2l-1)},t_{\sigma(2l)})~
\\ & ~~~~~  
 \sum_{m\leq k\leq\frac{n}{2}} \frac{(-1)^{k-m}}{2^{k-m}(k-m)!(n-2k)!} 
    \prod_{i=m+1}^{k} 
 g^{D}_{ab}(t_{\sigma(2i-1)},t_{\sigma(2i)})\prod_{j=2k+1}^{n}V(t_{\sigma(j)}) \end{split}\\
\begin{split}& =\int_a^b dt_{1}\ldots dt_{n}
\sum_{0\leq m \leq\frac{n}{2}}~\frac{(-1)^m}{2^m m!}~\prod_{l=1}^{m} 
\Delta^D_{ab}(t_{2l-1},t_{2l})~
\\ & ~~~~~  
 \sum_{\sigma \in S_{n-m}} ~ \sum_{m\leq k\leq\frac{n}{2}} \frac{(-1)^{k-m}}{2^{k-m}(k-m)!(n-2k)!} 
    \prod_{i=m+1}^{k} 
 g^{D}_{ab}(t_{\sigma(2i-1)},t_{\sigma(2i)})\prod_{j=2k+1}^{n}V(t_{\sigma(j)}) \end{split}\\
\begin{split}&=\sum_{0\leq m \leq\frac{n}{2}}~\frac{(-1)^m}{2^m m!}~\prod_{l=1}^{m} 
\int_a^b dt_{2l-1} dt_{2l} \Delta^D_{ab}(t_{2l-1},t_{2l})~
 \int_a^b dt_{m+1}\ldots dt_{n}
\\ & ~~~~~  
 \sum_{\sigma \in S_{n-m}} ~ \sum_{m\leq k\leq\frac{n}{2}} \frac{(-1)^{k-m}}{2^{k-m}(k-m)!(n-2k)!} 
    \prod_{i=m+1}^{k} 
 g^{D}_{ab}(t_{\sigma(2i-1)},t_{\sigma(2i)})\prod_{j=2k+1}^{n}V(t_{\sigma(j)}) \end{split}\\
\begin{split}& =\sum_{0\leq m \leq\frac{n}{2}}~ \frac{1}{m!} \left (-\half 
\int_a^b ds_{1} ds_{2} \Delta^D_{ab}(s_1,s_2) \right )^m ~~\times~\\ &
 \int_a^b dt_{1}\ldots dt_{n-2m} 
 \sum_{\sigma \in S_{n-m}}  \sum_{0\leq k\leq\frac{n}{2}-m} \frac{(-1)^{k}}{2^{k}(k)!(n-2m-2k)!} 
    \prod_{i=1}^{k} 
 g^{D}_{ab}(t_{\sigma(2i-1)},t_{\sigma(2i)})\prod_{j=2k+1}^{n-2m}V(t_{\sigma(j)})\end{split}\\
&= \sum_{0\leq m \leq\frac{n}{2}}~ \frac{1}{m!} \left (-\half 
\int_a^b ds_{1} ds_{2} \Delta^D_{ab}(s_1,s_2) \right )^m ~~ [V(a,b)^{n-2m}]_r
 \end{align}\end{subequations}
While the exponential form would automatically make the combinatorial
factors `work out', using this form makes it easier to ensure that
the integrand stays finite at every step, a crucial part of the 
proof.

\section{Proof of equation (\ref{eq.rtc.Q1-factorization-lemma})}
\label{lemma}

We wish to show that
\be
\int_{\Gamma_{\epsilon}^{ab}}d^{n}t\int_{\Gamma_{\epsilon}^{ab}}d^{2}s~f(\vec{t})\partial_{s_{2}}\left(g^{D}_{ab}(s_{1},s_{2})c(s_{2})\right)=\int_{\Gamma_{\epsilon}^{ab}}d^{n}t~d^{2}s~f(\vec{t})\partial_{s_{2}}\left(g^{D}_{ab}(s_{1},s_{2})c(s_{2})\right)
\ee
for any function $f(\vec{t})$ which is bounded on $(a,b)$.  
The difference between integrals over the two regions can be written in terms of three 
other other  integrals:
\begin{multline}\label{eq.rtc.dg-factorization-start}
\left(\int_{\Gamma_{\epsilon}^{ab}}d^{n}t\int_{\Gamma_{\epsilon}^{ab}}d^{2}s-\int_{\Gamma_{\epsilon}^{ab}}d^{n}t~d^{2}s\right)f(\vec{t})\partial_{s_{2}}\left(g^{D}_{ab}(s_{1},s_{2})c(s_{2})\right)\\
=\sum_{i=1}^{n}\int_{\Gamma_{\epsilon}^{ab}}d^{2}s\left(\int_{\Gamma_{\epsilon}^{ab}\cap|t_{i}-s_{1}|<\epsilon}d^{n}t+\int_{\Gamma_{\epsilon}^{ab}\cap|t_{i}-s_{2}|<\epsilon}d^{n}t\right)f(\vec{t})\partial_{s_{2}}\left(g^{D}_{ab}(s_{1},s_{2})c(s_{2})\right)\\
-\sum_{i=1}^{n}\int_{\Gamma_{\epsilon}^{ab}}d^{2}s\int_{\Gamma_{\epsilon}^{ab}\cap|t_{i}-s_{1}|<\epsilon\cap|t_{i}-s_{2}|<\epsilon}d^{n}t~f(\vec{t})\partial_{s_{2}}\left(g^{D}_{ab}(s_{1},s_{2})c(s_{2})\right)~.
\end{multline}
The first and second lines of the right hand side both vanish 
independently, so we will compute them separately, starting with the 
first line.  

Because the function $f$ is finite 
and is integrated over a region with area of order $\epsilon$, 
we notice that each of those integrals over $\vec{t}$ is $\epsilon$ 
times a finite function of one of the two remaining coordinates.  
Specifically, by defining
\be
F(s)=\frac{1}{\epsilon}\sum_{i=1}^{n}\int_{\Gamma_{\epsilon}^{ab}\cap|t_{i}-s|<\epsilon}d^{n}t~f(\vec{t})~,
\ee
the first line of \eq{eq.rtc.dg-factorization-start} is 
\begin{subequations}\be
\epsilon\int_{\Gamma_{\epsilon}^{ab}}d^{2}s~\partial_{s_{2}}\left(g^{D}_{ab}(s_{1},s_{2})c(s_{2})\right)\left(F(s_{1})+F(s_{2})\right)~.
\ee
We will not need to know the precise form of $F(s)$ so long as it and 
its derivative are finite.  With the full expression having an 
$\epsilon$ factor out front from the small area of the $t_{i}$ 
integral, we know that the finite part of $g^{D}_{ab}$ will not play 
any role, and we only need to consider the singular term.  
Integrating by parts, we have 
\begin{multline}
\epsilon\int_{a}^{b-\epsilon}ds\left(\frac{c(b)F(s)+c(b)F(b)}{(b-s)^{2}}-\frac{c(s+\epsilon)F(s)-c(s+\epsilon)F(s+\epsilon)}{\epsilon^{2}}\right)\\
+\epsilon\int_{a+\epsilon}^{b}\left(\frac{c(s-\epsilon)F(s)+c(s-\epsilon)F(s-\epsilon)}{\epsilon^{2}}-\frac{c(a)F(s)+c(a)F(a)}{(s-a)^{2}}\right)\\
-\epsilon\left(\int_{a}^{b-\epsilon}ds_{1}\int_{s_{1}+\epsilon}^{b}ds_{2}+\int_{a+\epsilon}^{b}ds_{1}\int_{a}^{s_{1}-\epsilon}ds_{2}\right)\frac{c(s_{2})F^{\prime}(s_{2})}{(s_{2}-s_{1})^{2}}~.
\end{multline}
The integrals with $\frac{1}{(b-s)^{2}}$ and $\frac{1}{(s-a)^{2}}$ 
can be done explicitly by Taylor expanding $F(s)$ about the 
appropriate endpoint.  The integrals with $\frac{1}{\epsilon^{2}}$ 
can be put over a common region by shifting the coordinate $s$ in one 
of them.  For the double integrals, we will Taylor expand the 
numerator about $s_{1}$ in order to perform the $s_{2}$ integral.
\begin{multline}
2cF(b)-2cF(a)+\int_{a+\epsilon}^{b}ds\frac{\left(c(s-\epsilon)-c(s)\right)\left(F(s-\epsilon)+F(s)\right)}{\epsilon}\\
-\epsilon\left(\int_{a}^{b-\epsilon}ds_{1}\int_{s_{1}+\epsilon}^{b}ds_{2}+\int_{a+\epsilon}^{b}ds_{1}\int_{a}^{s_{1}-\epsilon}ds_{2}\right)\left(\frac{cF^{\prime}(s_{1})}{(s_{2}-s_{1})^{2}}+\frac{\partial(cF^{\prime})(s_{1})}{s_{2}-s_{1}}+\ldots\right)
\end{multline}
Evaluating this further, we get 
\begin{gather}
\begin{split}2cF(b)-2cF(a)&-2\int_{a}^{b}ds~\partial c(s)F(s)\\
    &\quad-\epsilon\int_{a}^{b-\epsilon}ds\left(\frac{cF^{\prime}(s)}{\epsilon}-\frac{cF^{\prime}(s)}{b-s}\right)-\epsilon\int_{a+\epsilon}^{b}ds\left(\frac{cF^{\prime}(s)}{\epsilon}-\frac{cF^{\prime}(s)}{s-a}\right)\end{split}\\
=2cF(b)-2cF(a)-2\int_{a}^{b}ds~\partial c(s)F(s)-2\int_{a}^{b}ds~cF^{\prime}(s)+O(\epsilon\ln\epsilon)\\
=2cF(b)-2cF(a)-2\int_{a}^{b}ds~\partial_{s}\left(cF(s)\right)+O(\epsilon\ln\epsilon)~,
\end{gather}\end{subequations}
which goes to zero in the $\epsilon\rightarrow0$ limit.

Turning now to the last line in \eq{eq.rtc.dg-factorization-start}, 
where $t_{i}$ is close to both $s_{1}$ and $s_{2}$, we define 
\be
F_{2}(s_{1},s_{2})=\frac{1}{\epsilon}\sum_{j=1}^{n}\int_{\Gamma_{\epsilon}^{ab}\cap|t_{i}-s_{1}|<\epsilon\cap|t_{i}-s_{2}|<\epsilon}d^{n}t~f(\vec{t}),\quad F_{3}(s_{1},s_{2})=F_{2}(s_{1},s_{2})c(s_{2})~.
\ee
Both of these functions are finite for the same reasons as $F(s)$ 
above: they are finite functions integrated over a region with area 
proportional to $\epsilon$, and then divided by $\epsilon$.  The term 
we wish to evaluate is 
\begin{subequations}\begin{multline}
\epsilon\left(\int_{a+2\epsilon}^{b}ds_{1}\int_{s_{1}-2\epsilon}^{s_{1}-\epsilon}ds_{2}+\int_{a}^{b-2\epsilon}ds_{1}\int_{s_{1}+\epsilon}^{s_{1}+2\epsilon}+\int_{a+\epsilon}^{a+2\epsilon}ds_{1}\int_{a}^{s_{1}-\epsilon}ds_{2}+\int_{b-2\epsilon}^{b-\epsilon}ds_{1}\int_{s_{1}+\epsilon}^{b}ds_{2}\right)\\
F_{2}(s_{1},s_{2})\partial_{s_{2}}\left(g^{D}_{ab}(s_{1},s_{2})c(s_{2})\right)~.
\end{multline}
As with the other term, we will integrate this by parts.
\begin{multline}
\epsilon\int_{a+2\epsilon}^{b}ds\left(\frac{F_{3}(s,s-\epsilon}{\epsilon^{2}}-\frac{F_{3}(s,s-2\epsilon)}{4\epsilon^{2}}\right)+\epsilon\int_{a}^{b-2\epsilon}ds\left(\frac{F_{3}(s,s+2\epsilon)}{4\epsilon^{2}}-\frac{F_{3}(s,s+\epsilon)}{\epsilon^{2}}\right)\\
+\epsilon\int_{a+\epsilon}^{a+2\epsilon}ds\left(\frac{F_{3}(s,s-\epsilon)}{\epsilon^{2}}-\frac{F_{3}(s,a)}{(s-a)^{2}}\right)+\epsilon\int_{b-2\epsilon}^{b-\epsilon}ds\left(\frac{F_{3}(s,b)}{(b-s)^{2}}-\frac{F_{3}(s,s+\epsilon)}{\epsilon^{2}}\right)\\
-\epsilon\left(\int_{a+2\epsilon}^{b}ds_{1}\int_{s_{1}-2\epsilon}^{s_{1}-\epsilon}ds_{2}+\int_{a}^{b-2\epsilon}ds_{1}\int_{s_{1}+\epsilon}^{s_{1}+2\epsilon}+\int_{a+\epsilon}^{a+2\epsilon}ds_{1}\int_{a}^{s_{1}-\epsilon}ds_{2}+\int_{b-2\epsilon}^{b-\epsilon}ds_{1}\int_{s_{1}+\epsilon}^{b}ds_{2}\right)\\
\frac{\partial_{s_{2}}(F_{2}(s_{1},s_{2}))c(s_{2})}{(s_{2}-s_{1})^{2}}
\end{multline}
For the terms with a $\frac{1}{\epsilon^{2}}$ we will gather like 
denominators, shifting the integration variable when necessary to 
match intervals.  For the other single integrals, the functions 
$F_{3}(s,a)$ and $F_{3}(s,b)$ can be Taylor expanded about the 
endpoints $a$ and $b$ and only the first term will contribute, with 
the rest of the Taylor series giving at most terms of order 
$O(\epsilon\ln\epsilon)$.  For the double integrals, we will also Taylor 
expand $\partial_{s_{2}}F_{2}(s_{1},s_{2})c(s_{2})$ in $s_{2}$ about 
$s_{2}=s_{1}$ and again only the first term will contribute.  In 
addition, the last two double integrals will not contribute at all 
because the $s_{1}$ integrals there provide extra suppression.
\begin{multline}
\int_{a+2\epsilon}^{b}ds\frac{F_{3}(s-2\epsilon,s)-F_{3}(s,s-2\epsilon)}{4\epsilon}+\int_{a+\epsilon}^{b}ds\frac{F_{3}(s,s-\epsilon)-F_{3}(s-\epsilon,s)}{\epsilon}\\
+\epsilon F_{3}(b,b)\int_{b-2\epsilon}^{b-\epsilon}\frac{ds}{(b-s)^{2}}+\epsilon F_{3}(a,a)\int_{a+\epsilon}^{a+2\epsilon}\frac{ds}{(s-a)^{2}}\\
-\epsilon\left(\int_{a+2\epsilon}^{b}ds_{1}\int_{s_{1}-2\epsilon}^{s_{1}-\epsilon}ds_{2}+\int_{a}^{b-2\epsilon}ds_{1}\int_{s_{1}+\epsilon}^{s_{1}+2\epsilon}ds_{2}\right)\frac{\partial_{2}(F_{2}(s_{1},s_{1}))c(s_{1})}{(s_{1}-s_{2})^{2}}
\end{multline}
Here $\partial_{2}F_{2}$ is the derivative with respect to the second 
parameter, and $\partial_{1}$ will be with respect to the first.  Now 
we Taylor expand the numerators on the first line and evaluate an 
integral for everything else.
\begin{multline}
\frac{1}{2}\int_{a}^{b}ds\left(\partial_{1}-\partial_{2}\right)F_{3}(s,s)+\frac{F_{3}(b,b)-F_{3}(a,a)}{2}-\left(\int_{a+2\epsilon}^{b}ds+\int_{a}^{b-2\epsilon}ds\right)\frac{\partial_{2}F_{2}(s,s)c(s)}{2}
\end{multline}
In order to remove the middle term, we would like to change 
$\left(\partial_{1}-\partial_{2}\right)$ to 
$-\left(\partial_{1}+\partial_{2}\right)=-\partial_{s}$ in the first 
term, which we can do by adding an extra $\partial_{1}$ piece.
\begin{align}
-\half\int_{a}^{b}ds~\partial_{s}F_{3}(s,s)+\frac{F_{3}(b,b)-F_{3}(a,a)}{2}+\int_{a}^{b}ds~\partial_{1}F_{3}(s,s)-\int_{a}^{b}ds~\partial_{2}F_{2}(s,s)c(s)\hfill\\
\hfill=\int_{a}^{b}ds\left(\partial_{1}F_{2}(s,s)-\partial_{2}F_{2}(s,s)\right)c(s)
\end{align}
\end{subequations}
Now we look back at the definition of $F_{2}(s_{1},s_{2})$ and see 
that it is a symmetric function of its two parameters, so that the 
two derivatives are equal when acting on the line $s_{1}=s_{2}$.  We 
thus have zero for all of \eq{eq.rtc.dg-factorization-start}.


%

\bibliographystyle{JHEP}
\bibliography{refs}

\end{document}